\documentclass[traditabstract]{aa}
\usepackage{txfonts}
\usepackage{psfig}

\usepackage{graphicx}
\usepackage{natbib}
\bibpunct{(}{)}{;}{a}{}{,}
\usepackage{longtable}

\def\ie{{i.e.,~}}
\def\eg{{e.g.,~}}
\def\etal{{et al.~}}
\def\7c1756{{7C1756+6520}}
\def\7c1751{{7C1751+6809}}
\def\7c1805{{7C1805+6332}}

\def\deg{\ifmmode {^{\circ}}\else {$^\circ$}\fi}
\def\kms{\ifmmode {\rm\,km\,s^{-1}}\else
    ${\rm\,km\,s^{-1}}$\fi}
\def\ergcm2s{\ifmmode {\rm\,ergs\,cm^{-2}\,s^{-1}}\else
    ${\rm\,ergs\,cm^{-2}\,s^{-1}}$\fi}
\def\ergAcm2s{\ifmmode {\rm\,ergs\,cm^{-2}\,s^{-1}\,\AA^{-1}}\else
    ${\rm\,ergs\,cm^{-2}\,s^{-1}\,\AA^{-1}}$\fi}
\def\ergs{\ifmmode {\rm\,ergs\,s^{-1}}\else
    ${\rm\,ergs\,s^{-1}}$\fi}
\def\kmsMpc{\ifmmode {\rm\,km\,s^{-1}\,Mpc^{-1}}\else
    ${\rm\,km\,s^{-1}\,Mpc^{-1}}$\fi}

\def\spose#1{\hbox to 0pt{#1\hss}}
\def\simlt{\mathrel{\spose{\lower 3pt\hbox{$\mathchar"218$}}
     \raise 2.0pt\hbox{$\mathchar"13C$}}}
\def\simgt{\mathrel{\spose{\lower 3pt\hbox{$\mathchar"218$}}
     \raise 2.0pt\hbox{$\mathchar"13E$}}}

\begin{document}

\title{Large Scale Structures around Radio Galaxies at $z \sim 1.5$}

\author{Audrey Galametz\inst{1,2,3} 
\and Carlos De Breuck\inst{1} 
\and Jo\"{e}l Vernet\inst{1} 
\and Daniel Stern\inst{2} 
\and Alessandro Rettura\inst{4} \and \\
Chiara Marmo\inst{5} 
\and Alain Omont\inst{5} 
\and Mark Allen\inst{3}
\and Nick Seymour\inst{6,7}}


\institute{European Southern Observatory, Karl-Schwarzschild-Str. 2, D-85748 Garching, Germany [e-mail: {\tt agalamet@eso.org}]
\and Jet Propulsion Laboratory, California Institute of Technology, 4800 Oak Grove Dr., Pasadena, CA 91109, USA
\and Observatoire Astronomique de Strasbourg, 11 rue de l$'$Universit\'e, 67000 Strasbourg, France
\and Department of Physics and Astronomy, Johns Hopkins University, 3400 North Charles Street, Baltimore, MD 21218, USA
\and Institut d'Astrophysique de Paris, CNRS, Universit\'e Pierre et Marie Curie, Paris, France
\and Mullard Space Science Laboratory, UCL, Holmbury St Mary, Dorking, Surrey, RH5 6NT, UK
\and Spitzer Science Centre, Caltech, 1200 East California Boulevard, Pasadena, CA 91125, USA}

\abstract{We explore the environments of two radio galaxies at $z \sim 1.5$,
7C~1751+6809 and 7C~1756+6520, using deep optical and near-infrared
imaging.  Our data cover $15 \times 15 {\rm arcmin}^2$ fields around
the radio galaxies.  We develop and apply $BzK$ color criteria to
select cluster member candidates around the radio galaxies and find
no evidence of an overdensity of red galaxies within 2~Mpc of
7C~1751+6809.  In contrast, 7C~1756+6520 shows a significant
overdensity of red galaxies within $2$~Mpc of the radio galaxy, by a
factor of $3.1 \pm 0.8$ relative to the four MUSYC fields.  At small
separation ($r < 6\arcsec$), this radio galaxy also has one $z \sim
1.4$ evolved galaxy candidate, one $z \sim 1.4$ star-forming galaxy
candidate, and an AGN candidate (at indeterminate redshift).  This
is suggestive of several close-by companions.   Several concentrations
of red galaxies are also noticed in the full 7C~1756+6520 field,
forming a possible large-scale structure of evolved galaxies with
a NW-SE orientation.  We construct the color-magnitude diagram of
red galaxies found near ($r < 2$~Mpc) 7C~1756+6520, and find a clear
red sequence that is truncated at $K_s \sim 21.5$ (AB).  We also
find an overdensity of mid-IR selected AGN in the surroundings of
7C~1756+6520.  These results are suggestive of a proto-cluster at
high redshift.}


\maketitle

\section{Introduction}

Galaxy clusters are the most massive collapsed structures in the
universe, which make them an excellent tool for investigating
fundamental questions in astronomy.  For example, the evolution of
cluster number density depends sensitively upon $\Omega_0$, but
only weakly upon $\Lambda$ and the initial power spectrum
\citep[e.g.,][]{Eke1998}, and thus provides strong constraints on
cosmology.  Moderate-redshift clusters from well-defined samples
such as the {\it ROSAT} Deep Cluster Survey have been used to
constrain $\Omega_M$ and $\sigma_8$ \citep{Borgani2001}, while
\citet{Stern2009} use the ages of cluster ellipticals to constrain the equation of
state of dark energy. Distant
X-ray luminous clusters provide the best lever arm for such studies,
yet few have been found to date.  Because galaxy clusters supply
large numbers of galaxies at the same redshift, they also provide
unique resources to study the formation and evolution of galaxies.


Due to the sensitivity limits of current surveys, it remains challenging to 
identify a large sample of high redshift galaxy clusters using classical optical
and X-ray selection techniques. During the past decade, a new technique for detecting galaxy 
clusters at $z > 1$ has been to look at the immediate surroundings of high-redshift radio
galaxies \citep[HzRGs hereafter;~][]{Best2003, Venemans2005, Kodama2007}. 
Indeed, it is now well established that the host galaxies
of powerful radio sources are among the most massive galaxies in the universe \citep{Seymour2007}. 
At low redshift, radio galaxies are associated with giant ellipticals 
\citep[cD and gE galaxies; ][]{Matthews1964}, which are preferentially located in rich 
environments.  
Because they are so massive, radio galaxies are excellent signposts to pinpoint 
the densest regions of the universe out to very high redshifts \citep[e.g., ][]{Stern2003}. 
For example, this has been shown by the strong ($5\sigma$) overdensities of Ly$\alpha$ and 
H$\alpha$ emitters around HzRGs at $2.1 < z \leq 5.2$ \citep{Kurk2004A, Miley2004, 
Venemans2005, Venemans2007}, believed to be the progenitors of rich, local clusters. 
However, Ly$\alpha$ and H$\alpha$ emitters found in these environments
are small, faint, blue objects likely to be young star-forming galaxies 
and probably constitute a small fraction of both the number of cluster galaxies and 
the total mass of the cluster. 

Interestingly, overdensities at the highest redshifts often have a filamentary 
nature and extend beyond $\sim 2$~Mpc \citep{Croft2005}. \citet{Carilli2002}, 
in a detailed study of filaments in the field of PKS~1138-262, an HzRG at $z=2.1$, do not 
detect any extended X-ray emission, indicating that this structure has not yet had sufficient time to 
virialize.  However, \citet{Kurk2004B} show that some segregation has occured, with the 
H$\alpha$ emitters, tracing the more evolved population, more centrally concentrated 
than the younger Ly$\alpha$ emitters.  Therefore, the missing link between these proto-clusters 
and the classical X-ray confirmed clusters found out to $z \sim 1.4$ \citep[e.g., ][]{Mullis2005, Stanford2006} 
apparently occurs in the redshift range $1.4 < z \leq 2$. 
This redshift range is therefore particularly 
interesting for identifying clusters at a redshift beyond where the classical selection techniques are sensitive,
but at a redshift where clusters are already partly virialized with a core of older, massive galaxies in place.

In this paper, we present the study of the surroundings of two radio galaxies 
at $z \sim 1.5$. The next section describes the targets and the multi-wavelength 
data available for the two fields as well as how we derive the multi-band source catalogs.
The third section describes the color criteria we derive to select candidate massive cluster
members and the results of this selection. The properties of the cluster member candidates
are discussed in \S4. A study of the AGN candidates found in the two fields is also presented
in \S5. Section 6 describes possible close-by companions of one of our targeted radio galaxies, 7C~1756+6520. 
We discuss the results in \S7. We assume a $\Lambda$CDM cosmology with $H_0 = 70$ 
km s$^{-1}$ Mpc$^{-1}$, $\Omega_m = 0.3$ and $\Omega_{\Lambda} = 0.7$. The magnitudes are 
expressed in the AB photometric system unless stated otherwise. 

\section{The data}

\subsection{Target selection}

This work follows on the SHzRG project ({\em Spitzer} High-Redshift Radio 
Galaxy; Seymour et al. 2007), which was designed to study a representative sample 
of 70 radio galaxies at $1 < z \leq 5.2$ and their surroundings. SHzRG obtained 
rest-frame near- to mid-infrared photometry for this sample using all three 
cameras on board {\em Spitzer}. From this sample, we selected radio galaxies with
$z \sim 1.5$ for further study. From the seven such sources available in the 
Spring semester of the Northern hemisphere, we selected the two radio galaxies with
the most supporting data, 7C~1756+6520 ($z = 1.48$; R.A.: 17:57:05.44, Dec.: +65:19:53.11) 
and 7C~1751+6809 ($z = 1.54$; R.A.: 17:50:49.87, Dec.: +68:08:25.93). 
These two radio galaxies were first published in \citet{Lacy1992} as part of a sample
of $57$ radio sources selected at $38$~MHz. That paper presents high resolution
radio maps of both objects. Their redshifts were first presented in \citet{Lacy1999}.

\begin{table*}[!t]
\caption{Observations}
\label{targets}
\centering
\begin{tabular}{l c c l c c l l}
Instrument  & Pixel Scale  & Band & Wavelength & Bandwidth & FoV & Exp. Time$^{\mathrm{a}}$ & Seeing$^{\mathrm{b}}$  \\
                     & (arcsec/pix) &            &        (nm)       &      (nm)      &   (arcmin$^2$) &      (min)     & (arcsec) \\
\hline
Palomar/LFC     & $0.18$ & $B$   & $440$  & $100$ &  $588/566$ & $345$/$360$   & $\sim1$      \\
                -            &       -       & $z$   & $900$  & $180$ & $556/442$ & $60$/$135$   & $\sim1$       \\
CFHT/WIRCAM &  $0.3$  & $J$   & $1252$ & $158$ & $477/482$ & $182$/$219$  & $0.7$-$1$  \\
                 -            &      -        & $Ks$ & $2146$ & $325$ & $477/482$ &  $53$/$64$   & $0.7$-$1$  \\
{\it Spitzer}/IRAC & $0.61$ & IRAC1 & $3560$ & $750$ & $42/42$ & $2$/$2$ & $1.66$ \\
                  -            &      -       & IRAC2 & $4520$ & $1010$ & $42/42$ & $2$/$2$ & $1.72$ \\
                  -            &       -       & IRAC3 & $5730$ & $1420$ & $42/42$ & $2$/$2$ & $1.88$  \\
                  -            &       -       & IRAC4 & $7910$ & $2930$ & $42/42$ & $2$/$2$ & $1.98$  \\
\hline     
\end{tabular}            
\begin{list}{}{}
\item[$^{\mathrm{a}}$] FoV and exposure time for 7C~1751+6809 and 7C~1756+6520 respectively.
\item[$^{\mathrm{b}}$] Values of the mean FWHM for {\it Spitzer}/IRAC four bands.
\end{list}

\end{table*}

\subsection{Observations and data reduction}

\subsubsection{Palomar/LFC $B$-band data}

We imaged the two targets using the Bessel $B$-band filter 
of the Large Format Camera \citep[LFC;][]{Simcoe2000} on
the Palomar 5m Hale Telescope (see Table~1).  LFC is a prime focus, wide-field
optical imager with a well-sampled 24.6~arcmin diameter field, 
imaged by an array of six $2048 \times 4096$ pixel back-side
illuminated SITe CCDs. We observed each target for 6 hours in September 
2007. The nights were photometric with an average 1\arcsec seeing.

The LFC data were reduced using the {\tt MSCRED} package of IRAF, a suite 
of tasks designed to process multi-extension, large-format images from the new 
generation of optical cameras.  Processing followed standard optical procedures. 
A distortion correction was applied to each chip, first using the default solution 
for LFC, then matching the stars of the USNO-B1.0 catalog \citep{Monet2003}. 
The final stacked image was therefore astrometrized to the
USNO-B1.0 reference frame. For photometry, we calibrated the images using observations of 
standard stars from \citet{Landolt1992}. 
We then converted to AB magnitudes using: $B_{\rm AB} = B_{\rm Vega} - 0.1$. 
We derived the $3\sigma$ ($5\sigma$) detection limits using  
$1.5\arcsec$ diameter apertures uniformly distributed over the images and found limiting
magnitudes of $\sim 27.1$ ($\sim 26.6$).

\subsubsection{Palomar/LFC $z$-band data}

\begin{figure*}[!t] 
\begin{center} 
\includegraphics[width=10cm,angle=90]{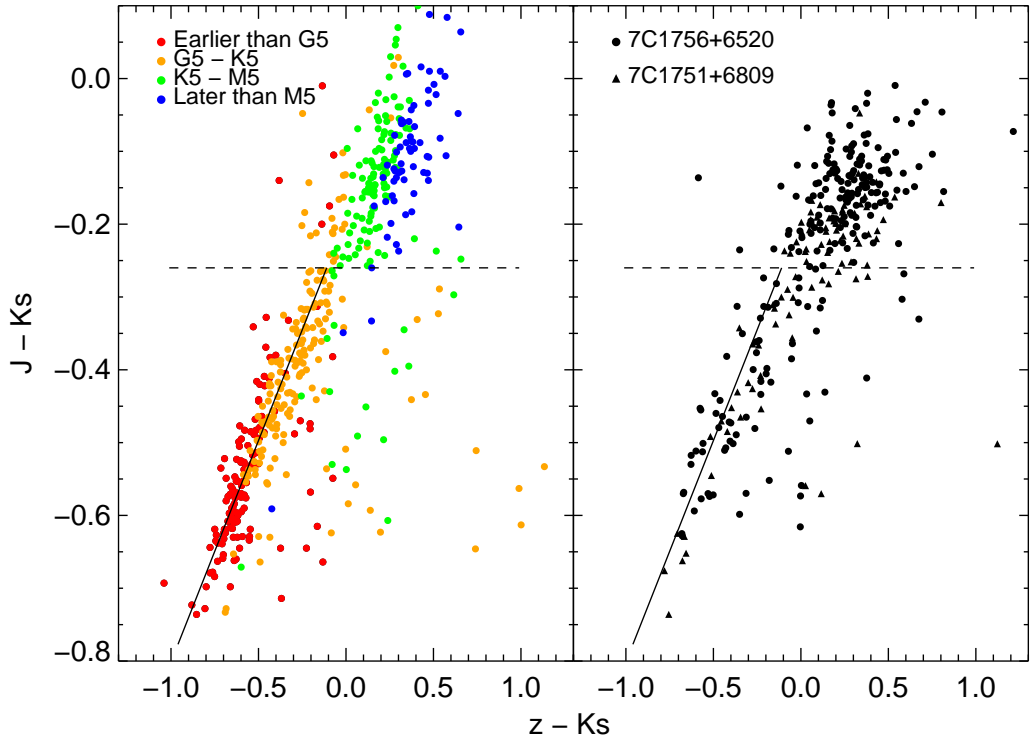}
\end{center}
\caption{Color-color diagrams for stars from the matched 2MASS/SDSS catalog (left) and for stars in the two 
radio galaxy fields (right). The spectral types of stars in the left panel were deduced from optical SDSS colors 
using \citet{Finlator2000} criteria. As illustrated by the solid line, stars with a spectral type earlier than K5 
($J - Ks < -0.26$) are well fit by an empirical color-color relation, $J - Ks = 0.61 (z - Ks) - 0.2$, which was used to
calibrate the optical images of the 7C fields
using 2MASS stars in each 7C field. The right panel shows the final color-color diagram of the 2MASS stars 
after calibration.}
\label{calibz}
\end{figure*}

We imaged the radio galaxy fields using the $z$-band filter
of Palomar/LFC (see Table~1). In February 2005, 
we observed 7C~1751+6809 for 60~min under photometric conditions.
In August 2005, we observed 7C~1756+6520 for 135~min but in non-photometric conditions.
The LFC data were reduced using the {\tt MSCRED} package of IRAF. The 
standard reduction process included an iterative removal of a $z$-band fringe 
pattern derived from the supersky flat as well as the same correction of distortion process
used for the $B$-band data. The final, stacked images were astrometrically registered to 
the USNO-B1.0 catalog. The FWHM of the final images is $\sim 1\farcs0$ for both fields.
Because these data were not all obtained in photometric conditions, nor were these fields covered 
by the Sloan Digital Sky Survey \citep[SDSS;][]{York2000}, photometric calibration of the $z$-band imaging
relied on empirically derived optical through near-IR color relations for Galactic stars. Matching a portion 
of SDSS imaging data with the Two Micron All Sky Survey \citep[2MASS;][]{Skrutskie1997}, 
\citet{Finlator2000} show that stars have a well-defined optical/near-infrared color locus, 
mainly determined by spectral type. We created a 2MASS/SDSS matched catalog of $530$ 
stars with $z < 18$ selected in three random extragalactic fields imaged by both SDSS and 2MASS. 
Following recent results from the SDSS 
collaboration\footnote[1]{See http://www.sdss.org/DR2/algorithms/fluxcal.html.},
SDSS $z$ band magnitudes are shifted by 0.02 relative to the AB system in the sense 
$z_{\rm AB} = z_{SDSS}+0.02$. We apply this systematic shift to the SDSS photometry and 
convert the $J$ and $K$ magnitudes from 2MASS to AB magnitudes using the following corrections: 
$J_{\rm AB} = J_{\rm Vega}+0.90$ and $K_{\rm AB} = K_{\rm Vega}+1.86$. 
Using the criteria defined in Finlator \etal (2000) and optical photometry from SDSS 
($g$, $r$ and $i$-band) to separate stars into spectral classes, we 
plot their location in a $J - K$ vs $z - K$ color-color diagram (Fig.~\ref{calibz}, left panel). 
Stars with spectral type K5 and earlier have $J - K < -0.26$ and a 
color-color relation well fit by a simple linear function: $J - K = 0.61 \times (z - K) - 0.2$.  Galaxies and cooler 
stars have redder $J - K$ colors.

Using 2MASS photometry, we identified stars with a spectral type earlier than K5
in our two radio galaxies fields assuming a $J - K \leq -0.26$ color. We selected $72$ and $40$ stars, 
respectively, for 7C~1756+6520 and 7C~1751+6809. 
Using the above color-color relation, we thus derived the $z$-band photometric zeropoints
for the Palomar data. The color-color diagram for stars in our fields is 
given in Fig.~\ref{calibz} (right panel). Measuring the dispersion of the empirical color-color 
relation, we estimate a $0.1$~mag uncertainty in the $z$-band photometric zeropoints. The 
$3\sigma$ ($5\sigma$) limiting magnitude determined from random $1.5\arcsec$ diameter apertures 
is $25.0$ ($24.5$) for 7C~1756+6520 and $24.8$ ($24.3$) for 7C~1751+6809.

\subsubsection{CFHT/WIRCAM data}

In order to sample the red side of the 4000 \AA~break at the redshift of the targets, the radio galaxies 
fields were observed in the $J$ and $Ks$ bands using the new Wide-field Infrared 
Camera \citep[WIRCAM;][]{Puget2004} of the Canada-France-Hawaii Telescope (CFHT; see Table~1). WIRCAM 
contains four $2048 \times 2048$ pixel HAWAII2-RG detectors with a gap of $45\arcsec$ between arrays, 
and covers a $20' \times 20'$ field of view (FoV) with a sampling of $0.3\arcsec$ per pixel. The imaging 
observations were obtained in April, May and July 2006 (Projects 06AF38 and 06AF99; P.I.~Omont). 
The seeing varied between $0.7$ and $1\arcsec$ during the observations and the nights were photometric. 

The WIRCAM data suffer from serious crosstalk, which echoes all bright objects in the $32$ amplifiers
of each chip. Although our HzRGs are at high Galactic latitude ($b > 30\deg$), our images contain
numerous bright stars due to the wide field of view of WIRCAM.
The crosstalk has different profiles and thus proves especially challenging to correct. 
Several techniques were attempted to correct crosstalk but none of them were fully satisfactory. 
For our total exposure time of approximately 3h30 in the $J$ band, the crosstalk is clearly visible for all 
objects brighter than magnitude 16. In the end, we processed the WIRCAM data without any 
crosstalk correction and instead flagged the most seriously affected regions (see \S2.3). The remaining 
processing followed standard near-infrared data reduction strategies. We subtracted the dark and 
performed flatfielding with a super flat created from science frames. The images were then sky 
subtracted and stacked using the reduction pipeline developed by the Terapix 
team\footnote[2]{http://terapix.iap.fr.} \citep{Marmo2007}. The images 
were photometrically calibrated to 2MASS $J$ and $K$ bands using $\sim60$ stars per field. 
The $3\sigma$ ($5\sigma$) limiting magnitudes determined from random 
$1.5\arcsec$ radius apertures in the $J$ and $Ks$ bands are $\sim 24.4$ ($\sim 23.9$) and $\sim 23.4$ 
($\sim 22.9$), respectively.

\subsubsection{Spitzer/IRAC data}

Observations with the {\em Spitzer} Infrared Array Camera \citep[IRAC;][]{Fazio2004} 
were performed as part of the GO-1 {\em Spitzer} program ``The Most Massive Galaxies at Every 
Epoch: A Comprehensive {\em Spitzer} Survey of High-Redshift Radio Galaxies'' (Seymour \etal 2007).  
These data consisted of four dithered $30$s exposures in each of the four IRAC channels (see Table~1). 
The size of the final IRAC mosaic is about $13' \times 7'$. Due to the configuration of the 
camera, only a $6.5' \times 6.5'$ region is covered with all four bands. The data were 
processed and mosaiced using the MOPEX package (Makovoz \& Khan 2005)\nocite{Makovoz2005} 
from the {\em Spitzer} Science Center and re-sampled by a factor of two. The final pixel 
scale is $0.61\arcsec$ (see Seymour \etal 2007 for further details on the {\em Spitzer} data and 
processing). The $5\sigma$ limiting magnitudes determined from random $1.5\arcsec$ radius 
apertures are $22.1$, $21.7$, $19.8$ and $19.7$ for the $3.6$, $4.5$, $5.8$ and $8.0\mu$m channels, respectively.

\subsection{Catalog extraction}

\begin{figure}
\includegraphics[width=9cm]{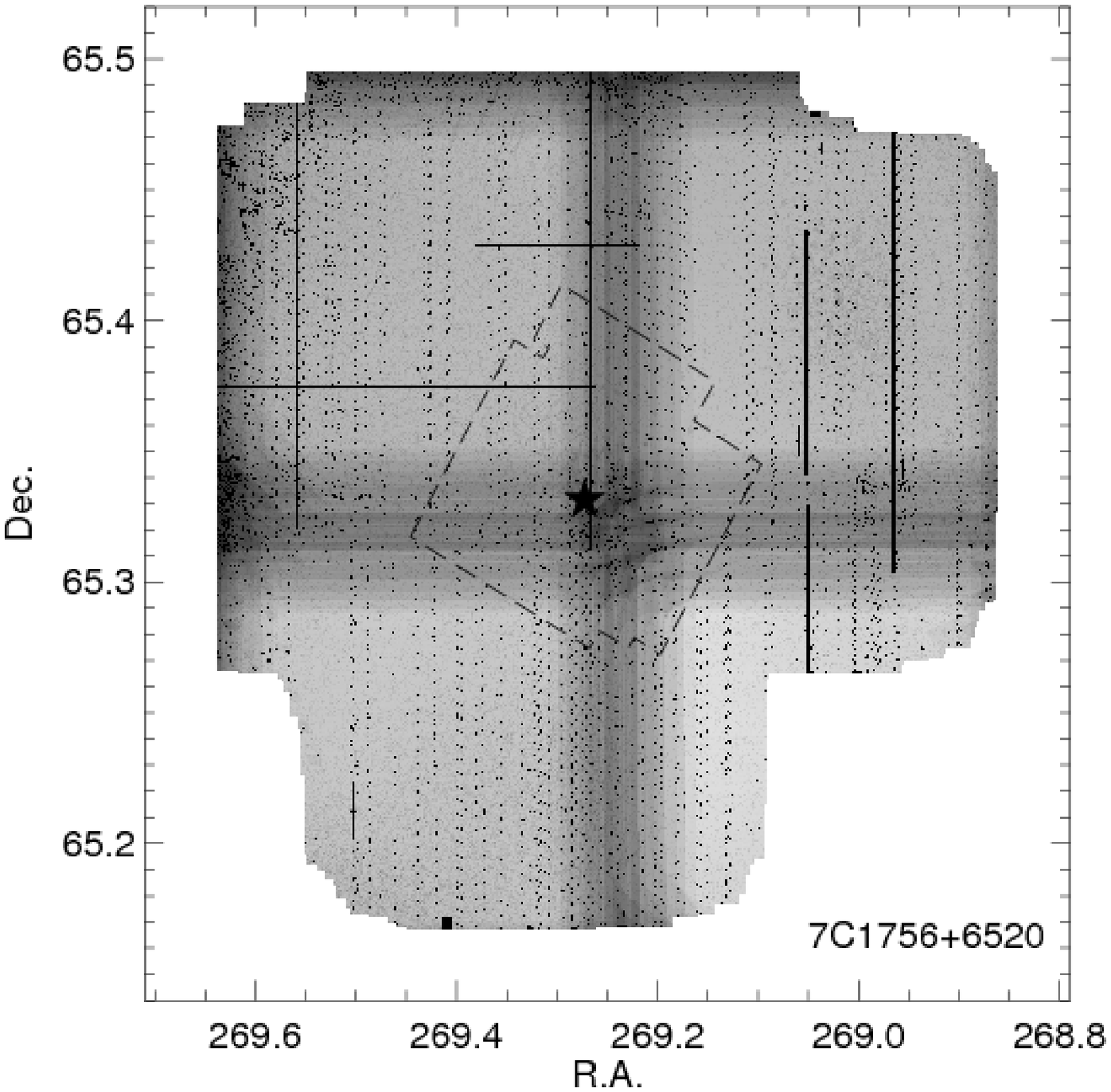} \\
\includegraphics[width=9cm]{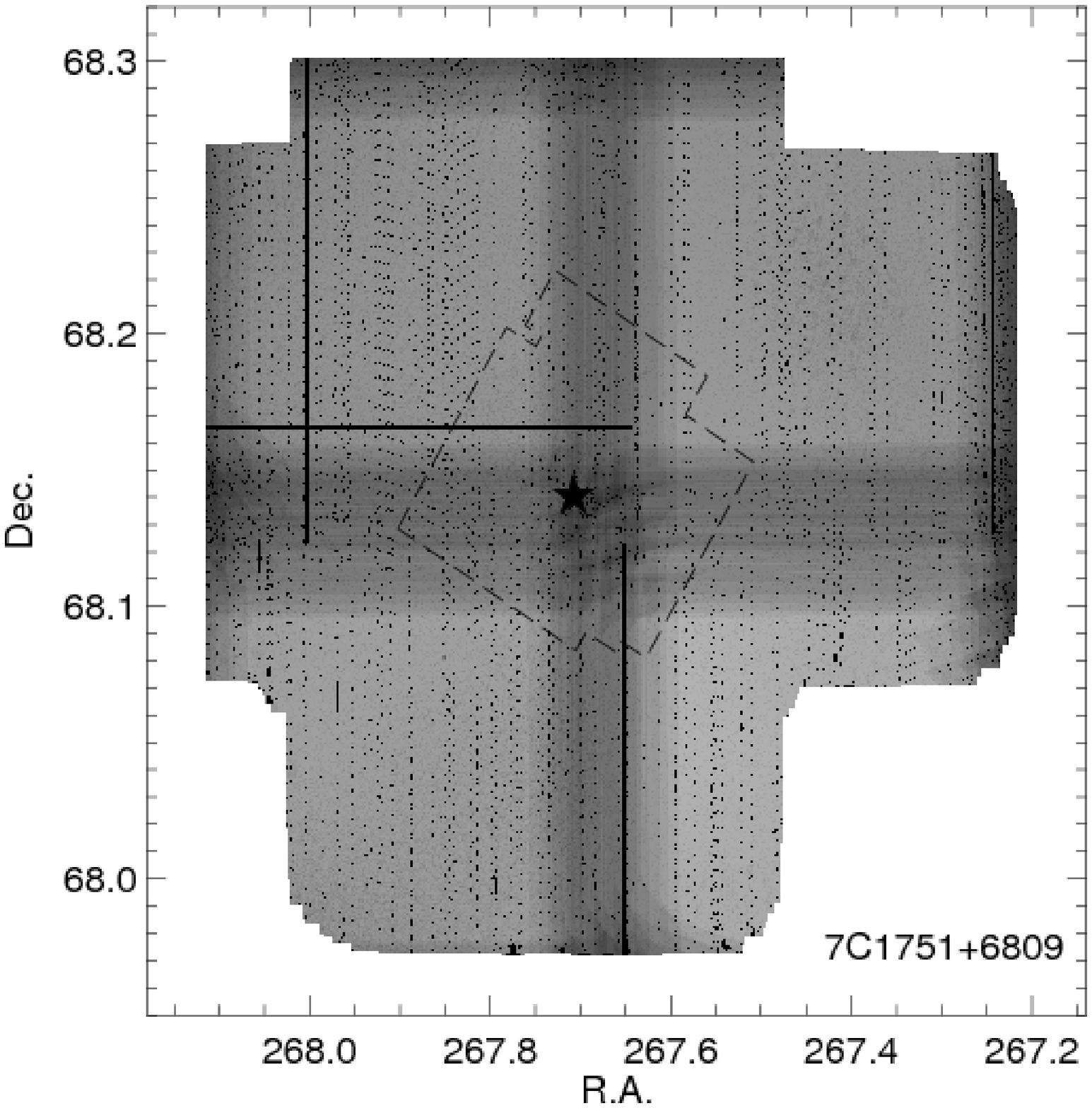} \\
\caption{Combined field covered by our $B$, $z$, $J$ and $Ks$-band data showing both the weight map
of the WIRCAM data ($J$) and the cross-talk flag map. We also flag regions contaminated by
bright stars. The dashed lines outline the regions covered by the four IRAC bands. The positions of the
HzRGs in the fields are indicated by stars.}
\label{weightmap}
\end{figure}

For the WIRCAM data, we identified crosstalk-affected pixels in the $J$ band image, the deeper of our
WIRCAM bands. A map was created to flag crosstalk contaminated pixels as well as the zones 
contaminated by bright star artifacts, which accounted for approximately $8$\% of the final mosaic 
pixels (see Fig.~\ref{weightmap}). The $J$ and $Ks$ images were smoothed to the $1\arcsec$ seeing of the 
$B$ and $z$ band data. We used SExtractor (Bertin \& Arnouts 1996) \nocite{Bertin1996} to 
extract source catalogs with SExtractor's dual mode for $J$ and $Ks$ using the unsmoothed images
for object detection and the smoothed one for photometry. For $B$, $z$, $J$ and $Ks$ bands, we derived 
colors using a fixed 2\farcs5 diameter aperture. For total magnitudes, we used the 
Kron automatic aperture photometry given by the SExtractor MAG\_AUTO parameter. 
All magnitudes were corrected for Galactic extinction using the dust maps of \citet{Schlegel1998} 
assuming the $R_V = A_V/E(B - V) = 3.1$ extinction law of \citet{Cardelli1989}.
Since both fields are at high Galactic latitude, their extinction maps are very uniform.
For both fields, the applied correction was $0.18$ in $B$-band, $0.06$ in $z$, $0.04$ in $J$ and
$0.02$ in $Ks$.

The point source function (PSF) of IRAC is well defined \citep{Lacy2005}, 
providing consistent and readily tabulated aperture corrections to determine 
total magnitudes from aperture photometry. For both magnitudes and colors, 
we chose an aperture of 2\arcsec5 diameter and corrected the measured flux 
by the corresponding multiplicative correction factors --- i.e., $1.68$, $1.81$, $2.04$ and $2.45$ 
for the $3.6$, $4.5$, $5.8$ and $8.0\mu$m channels, respectively.  

\begin{figure*}[!t]
\begin{tabular}{c c} 
\includegraphics[width=6.5cm,angle=90]{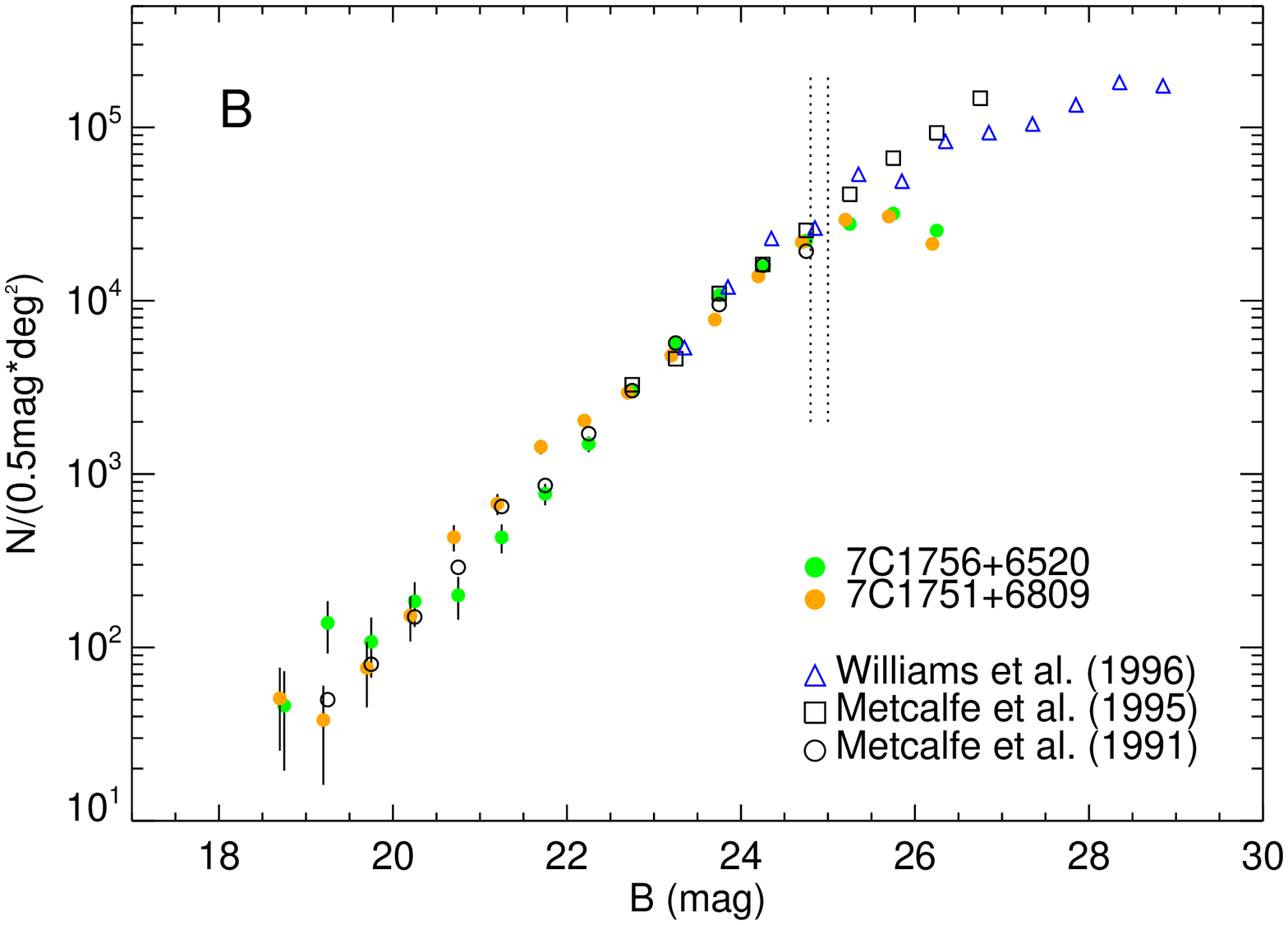} &
\includegraphics[width=6.5cm,angle=90]{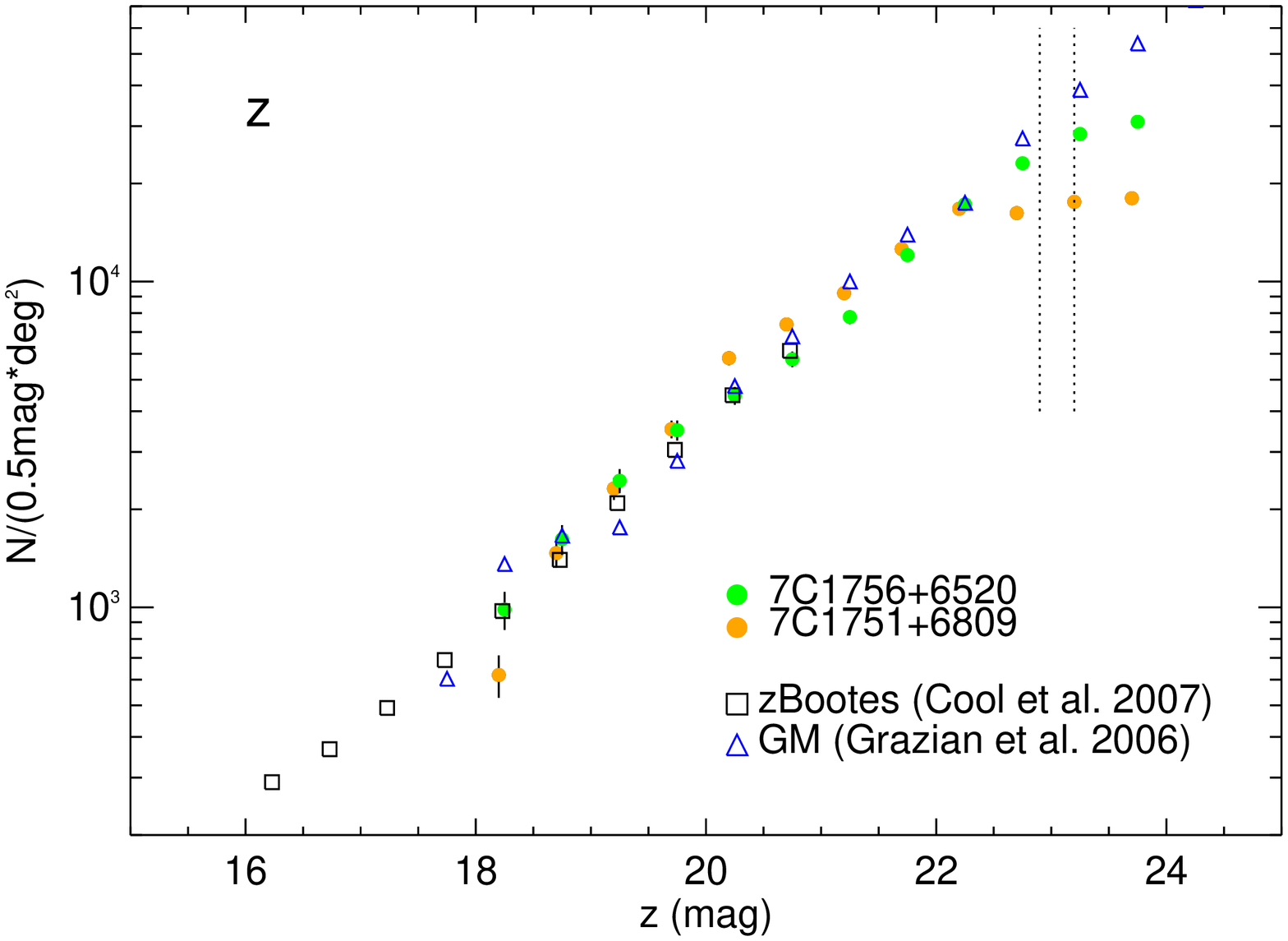} \\
\includegraphics[width=6.5cm,angle=90]{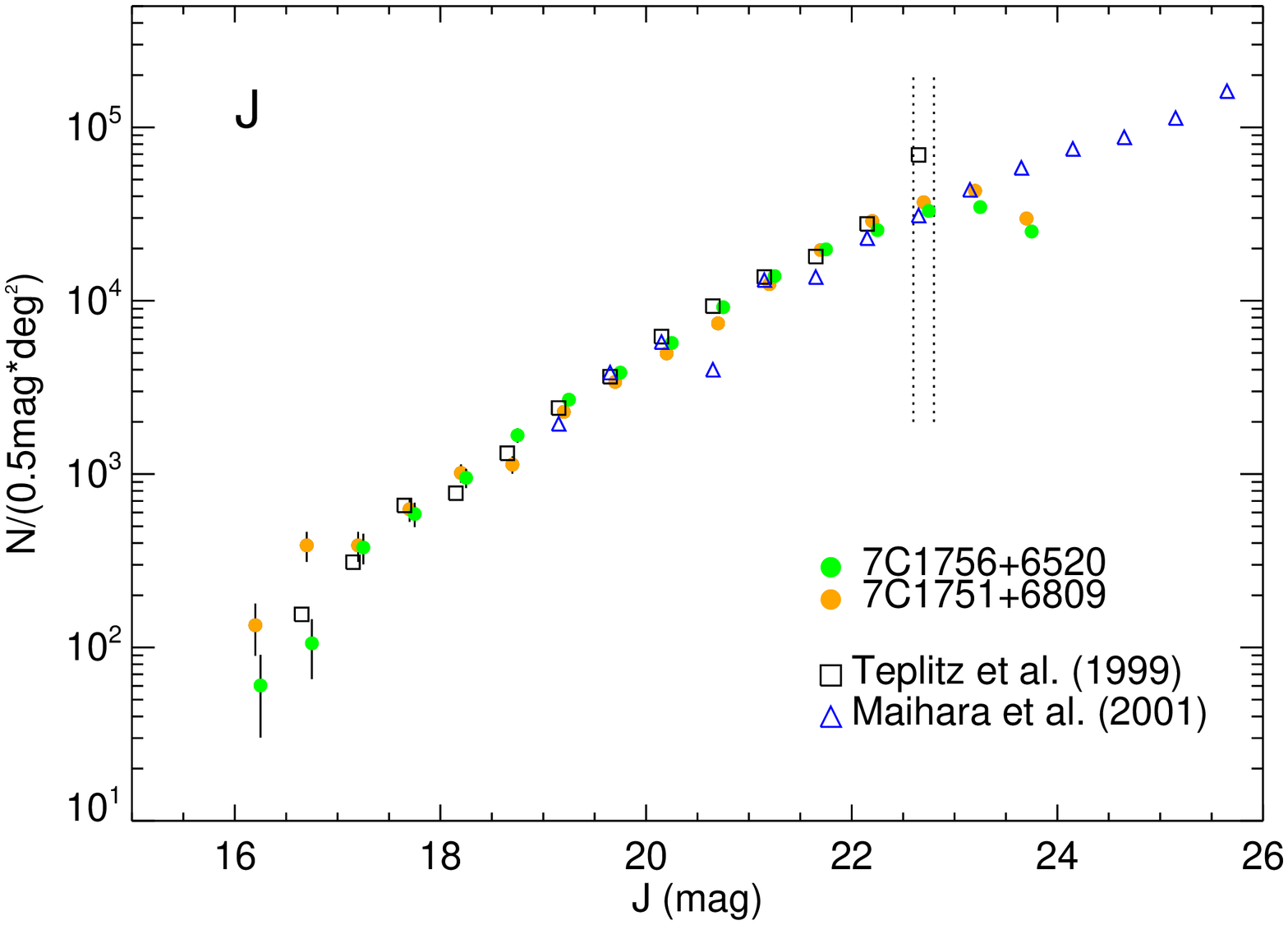} &
\includegraphics[width=6.5cm,angle=90]{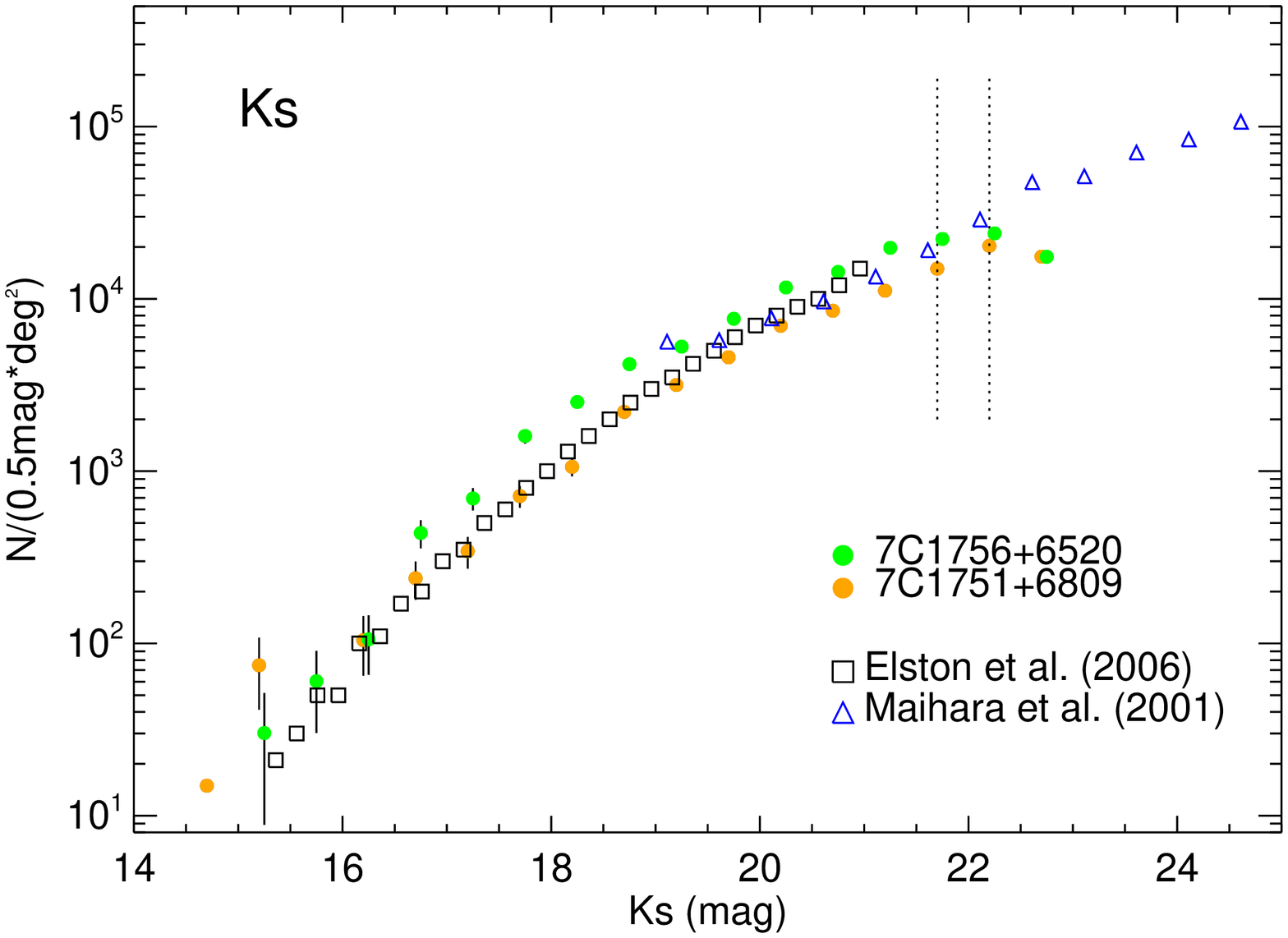} \\
\end{tabular}
\caption{Galaxy number counts for our $B$, $z$, $J$ and $Ks$ data. We use the stellar index 
determined by SExtractor (CLASS\_STAR) to separate galaxies from stars. No completeness 
correction was applied. The completeness limits of our images for elliptical and spiral galaxies are indicated 
by the two vertical dotted lines. We plot number counts from the literature 
(see legends for symbols; GM: GOODS-MUSIC). 
The counts in $B$, $z$ and $J$ are found in good agreement with the literature. However, the 
7C~1756+6520 field shows an excess of sources with $17 < Ks < 20.5$, the first evidence of 
an overdensity of very red objects in this field.}
\label{counts}
\end{figure*}

Combining all of these catalogs, we built a master catalog which provides multiwavelength
data for all sources detected in at least one of the eight 
bands observed. The final surface covered by $B$, $z$, $J$ and $Ks$ and not affected by 
the WIRCAM cross-talk is $\sim 0.1$ square degrees.
Fig.~\ref{counts} shows the galaxy number counts for the different bands compared 
with previous counts from the literature. The galaxies were first isolated from the stars 
based on SExtractor parameter CLASS\_STAR. The 1$\sigma$ error on the number counts
is overplotted on Fig.~\ref{counts}, assuming a Poisonnian error.
No incompleteness correction was applied to the counts. 

The galaxy counts determined from $B$, $J$ and $Ks$ were compared to previous works: 
\citet{Williams1996, Metcalfe1995, Metcalfe1991} for $B$, \citet{Maihara2001, Teplitz1999} for
$J$ and \citet{Elston2006, Maihara2001} for $Ks$. For the $z$-band, we derive number counts 
from zBo\"{o}tes \citep{Cool2007}, a $z$-band survey of the Bo\"{o}tes field that covers $7.62$
square degrees and reaches a $50\%$ completeness limit of $23.4$\footnote[3]{The final catalogs 
and images are available at http://archive.noao.edu/nsa/zbootes.html.}. We also derive $z$-band 
number counts from the GOODS-MUSIC catalog, a multiwavelength catalog of {\it Chandra} Deep 
Field South (CDFS) in the GOODS South field \citep[][see \S3.1 for details on this catalog]{Grazian2006B}.

The $B$, $z$ and $J$-band counts are found in good agreement with the literature. The $Ks$-band counts
are also found in agreement with previous studies for the field around 7C~1751+6809. The field
around 7C~1756+6520 however shows an excess of sources with $17 < Ks < 20.5$; at the faint limit, $Ks$ number
counts drop due to incompleteness. This overdensity is the first evidence of an overdensity of very
red objects around this radio galaxy. 

\subsection{Completeness}

In order to assess the completeness limit of our images, artificial galaxies of different types were added
to our images using the IRAF {\tt artdata} package ({\tt gallist} and {\tt mkobjects} routines). We first consider
the completeness limit for elliptical galaxies. For half magnitude intervals of brightness, we created 
catalogs of $5000$ elliptical galaxies which were randomly added to the $B$, $z$, $J$ and $Ks$ images, including 
Poisson noise. We adopted a de Vaucouleurs surface brightness law, a minimum galaxy axial ratio $b/a$ of $0.8$ 
and a maximum half flux radius of 1\farcs0. Running SExtractor with the same configuration files used for the 
unadulterated data, we determined the fraction of artificial sources detected.  For ellipticals, the derived 
90\% completeness limits for our images are $25.0$, $23.2$, $22.8$, and $22.2$ in the $B$, $z$, $J$ and $Ks$ 
bands, respectively. We then determined the completeness limit for spiral galaxies by creating catalogs of 
$5000$ spirals galaxies assuming an exponential disk surface brightness law with a minimum $b/a$ of $0.8$ and 
a maximum half flux radius of 1\farcs0. The derived $90$\% completeness limits are $24.8$, $22.9$, $22.6$ and 
$21.7$ in the $B$, $z$, $J$ and $Ks$ bands, respectively. As expected, the completeness limit for exponential
profile galaxies is slightly worse than for ellipticals due to the less compact nature of their morphologies.


\section{Candidate massive cluster members at $z \sim 1.5$}

We now consider the environments of 7C~1756+6520 and 7C~1751+6809. We first introduce a 
color criterion to select candidate cluster members based on the $BzK$ selection technique 
of \citet{Daddi2004}. We then discuss the selection of candidates selected using the full
multiwavelength master catalog (\S2.3) and finally present the results on the properties and 
clustering of these sources.

\subsection{Color selection of evolved galaxies at $z \sim 1.5$}

Substantial effort has gone into identifying color criteria to select galaxies and 
galaxy cluster members at high redshift. Selecting extremely red objects 
(EROs; $R-K \geq 4$), \citet{Stern2003} and \citet{Best2003} successfully 
identified evolved galaxy overdensities around HzRGs at $z \approx 1.1 - 1.6$. 
It has been shown that near-IR color criteria can be used to robustly identify 
passively evolving galaxies at $z \simgt 2$. These criteria are mainly based 
on the position of the 4000 \AA~break at a given redshift. Thus, the criterion 
$(J - Ks)_{\rm Vega} > 2.3$, which was first exploited by the FIRES team 
\citep{Franx2003}, is now well established and has been used to select cluster members at $z > 2$
\citep[Distant Red Galaxies, hereafter DRGs;][]{Kajisawa2006,Tanaka2007}. 
The galaxies selected by this criterion are mainly massive, evolved galaxies with old stellar
populations.  The goal of the current study is devise color criteria that are
optimized for identifying evolved galaxies at $z \simgt 1.4$,
sampling slightly higher redshifts than the ERO selection criteria,
but not as high redshift as the DRG or Lyman break selection criteria.

Based on the K20 survey \citep{Cimatti2002}, \citet{Daddi2004} proposed a simple two-color 
criterion based on $BzK$-band photometry for identifying galaxies at  
$1.4 \leq z \leq 2.5$ and classifying them as either star-forming galaxies, 
selected by $BzK \equiv (z - K) - (B - z) > -0.2$ (hereafter s$BzK$ galaxies) or passive 
evolving systems, selected by $BzK < -0.2 \cap (z - K) > 2.5$ (hereafter p$BzK$ galaxies). 
The $BzK$ selection is largely insensitive to dust extinction since $E(B-V)$ is
parallel to $BzK = -0.2$ criterion \citep{Daddi2004}. This two-color selection is 
therefore particularly efficient at isolating the red massive component of galaxy 
clusters at $z \geq 1.4$.

\begin{figure}[!t]
\psfig{file=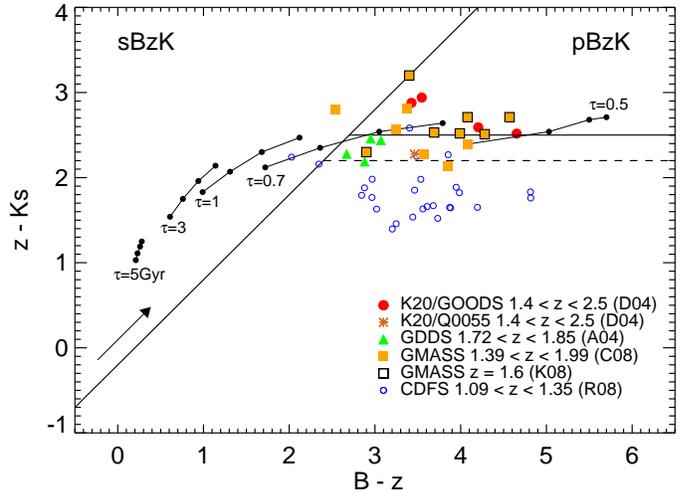,width=8.8cm,bbllx=50bp,bblly=35bp,bburx=540bp,bbury=400bp}
\caption{$BzK$ color-color plot of stellar population models at $z = 1.5$. 
The curves account for star formation histories with various $\tau$ models 
(respectively from left to right $\tau = 5, 3, 1, 0.7$ and $0.5$~Gyr).  For 
each model, black points indicate four different ages of the population 
($t = 2.5, 3, 3.5$ and $4$~Gyr). The black arrow indicates a dust extinction 
of E(B-V) = 0.2 as parameterized with the reddening curve of \citet{Cardelli1989}. 
The models are consistent with the $BzK$ criteria of Daddi \etal (2004; solid lines). 
However, the $BzK$ color selection for p$BzK$ galaxies is relatively strict and omits
some passive elliptical galaxies at $z \sim 1.5$. We plot the early-type galaxies with 
a spectroscopic redshift $1.4 < z < 2.5$ found in the literature (K20 - Daddi \etal 2004 - D04; 
GDDS - Abraham \etal 2004 - A04; GMASS - Cimatti \etal 2008 - C08; Kurk \etal 2008 - K08). We revise
the p$BzK$ criterion and adopt $z - Ks > 2.2 \cap  BzK < -0.2$ to select passive candidates at $z > 1.4$ (dashed line).
This criterion has been chosen as a compromise between selecting the majority of 
the passive galaxies candidates and avoiding contamination by lower redshift red objects. 
Contamination is expected to be small, though, as shown by the location of a sample of 
spectroscopically confirmed elliptical galaxies at lower redshift ($1.09 < z < 1.35$ - 
CDFS - Rettura \etal 2008 - R08) with only $15$\% of them selected by our extended $BzK$ criteria.}
\label{z14}
\end{figure}

We consider first the colors of different stellar populations at $z\sim1.5$ 
obtained from the Bruzual \& Charlot (2003)\nocite{Bruzual2003} models (Fig.~\ref{z14}). 
The different curves show the various dust-free $\tau$ models 
predictions (from left to right, $\tau = 5, 3, 1, 0.7, 0.5$ and $0.1$~Gyr), assuming solar
metallicity and a Salpeter (1955) initial mass function. For each model, 
four different population ages are indicated ($t = 2.5, 3, 3.5$ and $4$~Gyr). As 
previously stated, the $BzK$ criterion is relatively insensitive to dust extinction since the reddening 
vector for an extinction of $E(B-V) = 0.2$ is almost parallel to the $BzK = -0.2$ line (see 
black arrow in Fig.~\ref{z14}).

The model colors are consistent with the $BzK$ selection criterion, 
with models covering first the s$BzK$ zone and then the p$BzK$ zone of
the $BzK$ diagram as $\tau$ decreases. We note however that the criterion is most likely missing early-type 
galaxies at $z > 1.4$, in particular those with the youngest stellar populations (models with small 
$\tau$ and large $t$ values). We overplot a sample of the (rare) examples of early-type galaxies at $1.4 < z < 2$ and 
spectroscopically confirmed in the literature. \citet{Daddi2004} report five high redshift 
early-type galaxies from the K20 survey, classified as such on the basis of continuum breaks 
and absorption lines in their spectra. Four are in the GOODS area and one is in the Q0055 area.
\citet{Cimatti2008} used the Galaxy Mass Assembly ultra-deep Spectroscopic Survey \citep[GMASS; ][]{Kurk2008A} 
to find passive galaxies at $z > 1.4$. They used the UV properties of passive galaxies and 
derived a color index of the UV continuum for galaxies with spectroscopic redshift $z >1$ 
(see Cimatti \etal 2008 for details). Thirteen passively evolving galaxies at $1.390 < z < 1.981$ 
were found in GMASS, seven of which are members of an overdensity at $z \sim 1.6$ 
\citep{Kurk2008B}. The Gemini Deep Deep Survey \citep[GDDS; ][]{Abraham2004} obtained spectroscopy
for $309$ objects attempting to target galaxies in the `redshift desert' 
($1 < z < 2$)\footnote[4]{The GDDS catalog is publicly available at
http://lcirs.ociw.edu/public/GDDSSummary-dist.txt. Targeted magnitudes are in the Vega system. 
We convert from Vega to the AB photometric system using the corrections adopted earlier in this paper.}. 
Fifty of these sources have $BzK$ photometry (SA12 and SA15 fields) and $z \ge 1.4$, of which five have 
$BzK < -0.2$. One of these sources is at $z > 2$ and has a 
$z - K$ which is far too blue to be considered as a passively evolving galaxy ($z - K < 1$). We therefore 
find only four strong candidates for passively evolving galaxies in GDDS. The location of all these passive galaxies 
in the $BzK$ diagram is given in Fig.~\ref{z14}. Nine out of $22$ are found to have $z - K < 2.5$. 
We thus confirm what we had already suspected from the models, i.e. the $BzK$ criterion for the p$BzK$ selection 
is missing a significant fraction ($\sim40$\%) of old galaxies at $z > 1.4$.

We revise the p$BzK$ criterion and adopt $z - Ks > 2.2$ rather than $z - Ks > 2.5$, coupled 
with $BzK < -0.2$, to select passively evolving galaxies at $z > 1.4$ (hereafter p$BzK$* galaxies). 
This color cut has been chosen as a compromise between following the elliptical model color 
predictions as well as selecting the majority ($91\%$) of spectroscopically confirmed passive 
systems at $z > 1.4$ to date and minimizing contamination from very red galaxies at lower 
redshift. \citet{Rettura2008} study a sample of $27$ early-type galaxies found in the CDFS 
with $1.09 < z < 1.35$. Out of $27$, only four (all with $z >1.3$) are selected with our 
extended $BzK$ criteria (see Fig.~\ref{z14}, open circles; $BzK$ photometry from A.Rettura, 
private communication). We are therefore confident that the contamination of lower redshift 
red objects is small. Indeed, since the 4000\AA~break is at the red end of the $z$-band at 
$z \sim 1.4$, the $z - Ks$ color increases rapidly with redshift for $z\sim1.4$ making this 
simple color criteria an efficient redshift indicator, especially for passive systems. 

\citet{Grazian2006B} presents the GOODS MUlticolor Southern Infrared Catalog 
(GOODS-MUSIC), a multiwavelength catalog of the GOODS South field, combining imaging 
ACS (optical), VLT (near-infrared), and {\em Spitzer} (mid-infrared) data with available 
spectroscopic data. \citet{Grazian2006B} applied a photometric redshift ($z_{\rm phot}$) 
code to this multiwavelength dataset\footnote[5]{The full catalog, including photometric redshifts, 
is publicly available at http://lbc.mporzio.astro.it/goods/goods.php.}. For this study, we used an updated 
version of the GOODS-MUSIC catalog (version $2$) recently presented in \citet{Santini2009}. 
The new catalog contains, among other things, additional spectroscopic redshift and new
MIPS $24\mu$m photometry. The total area covered by the GOODS-MUSIC catalog is $143.2$ 
sq. arcmin. We check the revised $BzK$ selection technique using the photometric redshifts ($z_{phot}$ 
hereafter) of the p$BzK$ ($65$), p$BzK$* ($116$) and s$BzK$ ($4727$) galaxies found 
in the GOODS-MUSIC catalog. Of the p$BzK$ galaxies, $56$\% ($78$\%) are found with 
$z_{phot}>1.4$ ($z_{phot}>1.2$). The corresponding percentages are $49$\% ($74$\%) for
the p$BzK$* galaxies and $82$\% ($88$\%) of the s$BzK$ galaxies. Using the same 
photometric redshift code as used for GOODS-MUSIC $z_{phot}$, 
\citet{Grazian2006A} estimate an accuracy of $\sigma_{z} = 0.05\times(1+z)$ for red 
galaxies ($J-K>0.7$) and $\sigma_{z} = 0.03\times(1+z)$ 
for their full sample. Fig.~2 of the same paper shows that, at all redshifts, the photometric redshifts 
systematically underestimate the spectroscopic redshifts. Similar results 
were also found in \citet{Mobasher2004} whose photometric redshifts in the GOODS Southern 
Field at $z>1.3$ were also underestimated. The percentages presented above are thus likely
lower limits. The revised $BzK$ selection is therefore very efficient at isolating red galaxies
at $z>1.4$, with some inevitable contamination by lower redshift reddened galaxies.


\subsection{Candidate cluster members}

The combination of filters used during the observations were checked for consistency with the 
one used by \citet{Daddi2004}. Comparing the shape of the filter transmission curves, we deduce 
that the $B$-band filters are equivalent. The $z$-band filter of Palomar/LFC is consistent with the 
Gunn $z$-band of VLT/FORS1 though it is shorter at long wavelength by $\sim 400$\AA. Finally, the 
CFHT/WIRCAM $Ks$-band filter is slightly more extended at bluer wavelength (by $\sim 300$\AA)
compared to the one used at VLT/ISAAC by \citet{Daddi2004}. We use a library of galaxy 
templates generated with P\'EGASE2 (Projet d'\'Etude des Galaxies par 
Synth\`ese \'Evolutive; Fioc \& Rocca Volmerange 1997)\nocite{Fioc1997} 
and compare the colors obtained with the different filter sets. We conclude that the 
correction to the $B - z$ color is negligible, especially for galaxies at intermediate to 
high redshift ($z \geq 1.4$ - less than 0.02) and the correction to $z - Ks$ color is not systematic 
(\ie depending on the galaxy type and age) and are generally smaller than the calibration error 
of our $z$-band photometry (e.g.~$\leq$ $0.1$~mag on average; see \S2.2.2).

We next verify that the depth of our data is sufficient to select passively evolving systems at $z > 1.4$. 
The magnitudes of early-type galaxies at $z > 1.4$ and confirmed spectroscopically 
are, unfortunately, rarely given in the literature. Furthermore, the selection of such objects 
itself is strongly biased to the brightest objects. 
We look at the expected magnitudes of $pBzK$* galaxies in the GOODS-MUSIC catalog (see \S3.1). 
$57$ objects have $z_{\rm phot} > 1.4$ and $BzK$ magnitudes 
fitting the p$BzK$* criteria (hereafter the ``GM sample''). These sources have $24.8 < B < 29.7$ 
($\langle B \rangle \sim 28.2$), $22.2 < z < 26.2$ ($\langle z \rangle \sim 24.4$) and 
$19.7 < Ks < 24.4$ ($\langle Ks \rangle \sim 21.8$).
Considering the $3\sigma$ limits of our imaging, our $Ks$ data would detect $98$\% of the GM sample, our $z$ data
would detect $75$\% of the GM sample, but our $B$ data would detect only $11$\% of the GM sample.
Therefore, we treat the $z$-band as the limiting band for this work and consider sources with
upper limits in $B$. However, at a given $B$ magnitude, fainter 
objects in $z$ will have a bluer $B-z$ color, corresponding to bluer objects. We are therefore confident that the majority
of very red passive members of the clusters will be selected in our dataset.

\begin{figure}
\includegraphics[width=7.8cm,angle=0, bb=35 35 480 470]{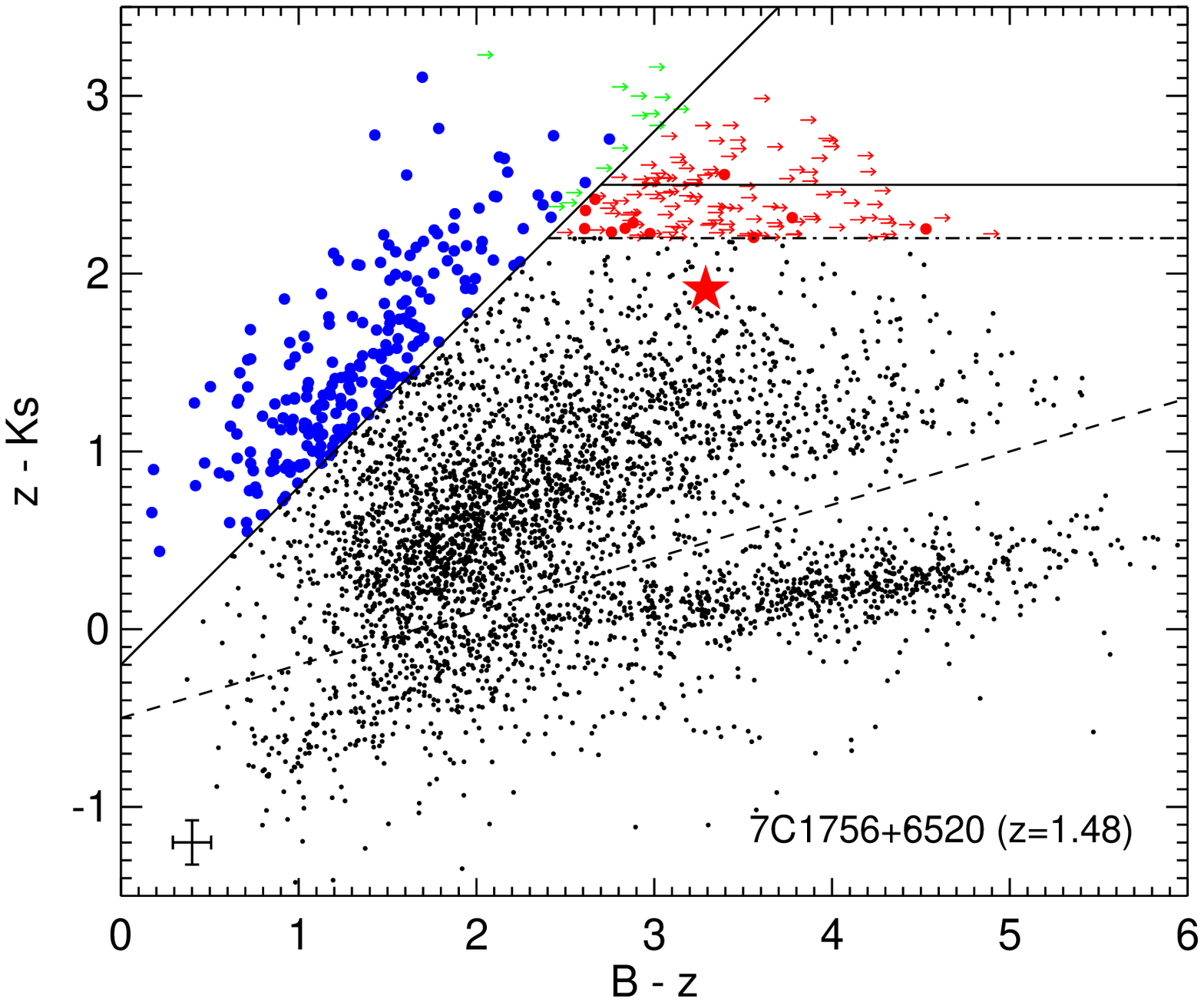} 
\includegraphics[width=7.8cm,angle=0, bb=35 35 480 470]{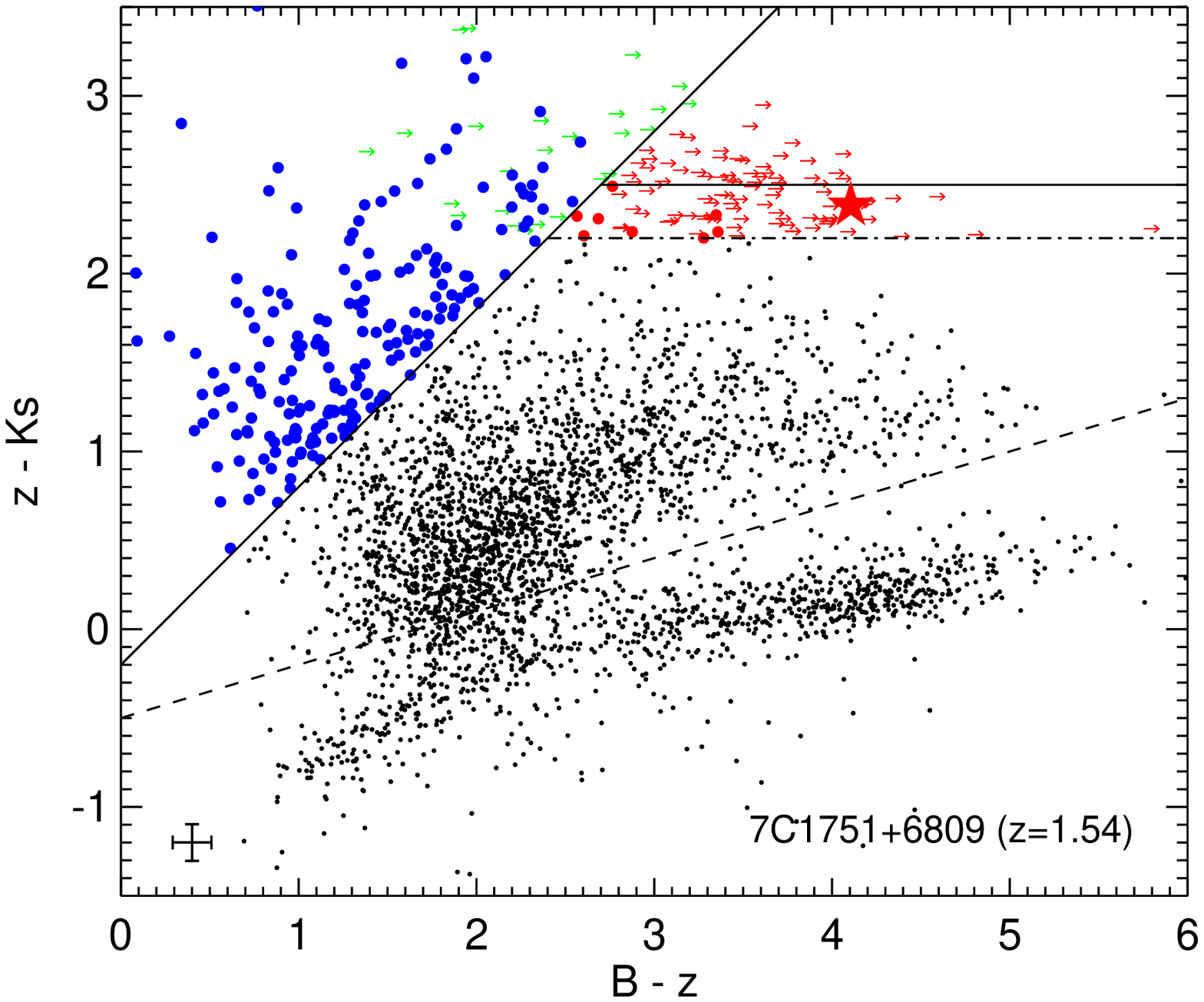} 
\caption{Color-color $BzK$ diagram of 7C~1756+6520 at $z = 1.48$ (top) and 7C~1751+6809 at $z = 1.54$ (bottom).
The s$BzK$ and p$BzK$ selection regions defined by \citet{Daddi2004} are shown by the solid lines. Our new 
p$BzK$* selection is shown by the dot-dashed line. The dashed line separates stars and galaxies.
All sources with a $3\sigma$ detection in $B$, $z$ and $Ks$ are plotted. Also plotted 
are the sources with a $z-Ks > 2.2$ but no (or $< 3\sigma$) detection in the $B$-band (arrows). In order to 
place those sources in the plot, we assign them the $3\sigma$ detection limit for the $B$ magnitude ($B = 27.1$).
Typical uncertainties on colors are indicated in the lower left corner of each plot. HzRGs are marked 
as red stars (7C~1751+6809 has only an upper limit in the $B$-band).}
\label{BzK}
\end{figure}


\begin{table*}
\caption{Surface densities of $BzK$ galaxies  (to the completeness limits; degrees$^{-2}$)}
\label{table2}
\centering
\begin{tabular}{l || c | cccc | c || cc | cc }
Field & GOODS & \multicolumn{4}{c|}{MUSYC} & All & \multicolumn{2}{c|}{7C~1756+6520} & \multicolumn{2}{c}{7C~1751+6809} \\
 & MUSIC & ECDFS & EHDFS & SDSS1030 & CW1255  & MUSYC & Full & $r<2$~Mpc & Full &$r<2$~Mpc \\
\hline
Area$^{\mathrm{a}}$ & $143.2$ & $969.6$ & $201.1$ & $106.7$ & $101.9$ & $1379.3$ & $333.4$ & $47.9$ & $336.2$ & $47.7$ \\
(1) & (2) & (3) & (4) & (5) & (6) & (7) & (8) & (9) & (10) & (11) \\
\hline
\hline
p$BzK$ & $130 \pm 60$ & $70 \pm 20$ & $70 \pm 40$ &	 $170 \pm 80$ & $35 \pm 35$ & $70 \pm 10$ & $190 \pm 45$ & $330 \pm 160$ & $140 \pm 40$ & no detection\\
\hline
p$BzK$* & $430 \pm 100$ & $370 \pm 40$ & $320 \pm 80$ & $640 \pm 150$ & $210 \pm 90$ & $370 \pm 30$ & $620 \pm 80$ & $1160 \pm 295$ & $510 \pm 70$ & $250 \pm 140$ \\
\hline
s$BzK$ & $500 \pm 110$ & $505 \pm 40$ & $480 \pm 90$ & $300 \pm 100$ & $490 \pm 130$ & $485 \pm 40$ & $490 \pm 70$ & $990 \pm 270$ & $350 \pm 60$ & $500 \pm 190$ \\
\hline
\end{tabular}
\begin{list}{}{}
\item[$^{\mathrm{a}}$] Area are given in square arcmin.
\end{list}
\end{table*}

We select the s$BzK$, p$BzK$ and p$BzK$* galaxies around 7C~1756+6520 and 
7C~1751+6809 using our multi-wavelength catalog. The coordinates and $B$, $z$, $J$ and $Ks$ 
magnitudes of the p$BzK$* galaxies are given in Tables 5 and 6 for 7C~1756+6520 and 
7C~1751+6809 respectively. We assume Poisson errors for 
source density determinations. Fig.~\ref{BzK} shows the $BzK$ color diagram of 
all the objects with a $3\sigma$ detection in $B$, $z$ and $Ks$. We also plot the sources 
that have a $z - Ks > 2.2$ but no (or $< 3\sigma$) detection in the $B$-band (arrows). 
In order to place those sources in Fig.~\ref{BzK}, we assign them the $3\sigma$ detection limit 
for the $B$ magnitude ($B = 27.1$). For 7C~1756+6520 (7C~1751+6809), we found $129$ ($106$) 
p$BzK$* galaxies including $42$ ($42$) p$BzK$ galaxies (with a $3\sigma$ detection in $z$ and $Ks$). This gives 
a surface density of $0.39 \pm 0.03$ ($0.32 \pm 0.03$) arcmin$^{-2}$ for 
p$BzK$* galaxies and $0.13 \pm 0.02$ ($0.12 \pm 0.02$) arcmin$^{-2}$ for p$BzK$ galaxies. 
We extract the star-forming candidates with a $3\sigma$ detection in $B$, $z$ and $Ks$ and 
found $218$ ($200$) s$BzK$ galaxies, \ie a surface density of $0.65 \pm 0.05$ 
($0.59 \pm 0.04$) arcmin$^{-2}$. $14$ ($26$) sources have $z- Ks > 2.2$ and no 
(or $< 3\sigma$) detection in $B$ (green arrows) and can not be reliably distinguished as 
star-forming or passive systems.

Considering the $J-Ks$ color of the $BzK$ sources, there is a clear difference between 
s$BzK$ and p$BzK$ galaxies, with p$BzK$ galaxies having  a redder and narrower 
distribution centered around $\langle J-Ks \rangle \simeq 1.07$ ($0.93$) for the 
7C~1756+6520 (7C~1751+6809) field. s$BzK$ galaxies have a 
$\langle J-Ks \rangle \simeq 0.71$ ($0.57$). We note that the p$BzK$* galaxies found around 
7C~1756+6520 are, on average, redder that those found in the 7C~1751+6809 field.

\subsection{Surface density of $BzK$-selected galaxies}

We now compare the density found in our HzRG fields to blank fields. 
\citet{Grazian2007} study the properties of various classes of high redshift 
galaxies, including p$BzK$ and s$BzK$ sample in the GOODS-MUSIC 
sample. They compare their number densities of s$BzK$ and p$BzK$ 
galaxies with the literature \citep{Daddi2004, Kong2006, Reddy2006} and 
conclude that the GOODS-South field is representative of the distant universe. 
We therefore use the GOODS-South as a first comparison field for our HzRG field. 
We cut the GOODS-MUSIC catalog at the same completeness limit as our data, 
\ie we select p$BzK$ and p$BzK$* galaxies to our $90$\% completeness limits of
$Ks < 22.2$ and $z < 23.2$, and we select s$BzK$ galaxies to $Ks < 21.7$ 
and $z < 22.9$. The number densities of s$BzK$, p$BzK$ and p$BzK$* 
galaxies in the GOODS-MUSIC catalog are given in Table 2 (column $2$)
assuming Poisson errors on the numbers. The corresponding densities in our 
two fields (corrected for incompleteness) are given in columns $8$ and $10$. 

We also compare our results to the MUSYC survey. The MUSYC survey 
\citep{Gawiser2006, Quadri2007} consists of four fields: an extended 
Hubble Deep Field South (E-HDFS), an extended Chandra Deep Field 
South (E-CDFS) and two fields called SDSS~1030 and CW~1255. 
We note that the region covered by GOODS-MUSIC is included in the E-CDFS.
Optical and near-infrared imaging of $30\arcmin \times 
30\arcmin$ were obtained for all the fields. Deeper near-infared imaging 
of $10\arcmin \times 10\arcmin$ were obtained for subfields of SDSS~1030 
and CW~1255 as well as for two adjacent subfields of  E-HDFS (resulting 
in a deeper subfield of $20\arcmin \times 10\arcmin$ for E-HDFS). 
The MUSYC team did not obtain additional data for E-CDFS since the 
region had already been observed extensively by the GOODS team. All 
images and photometric catalogs are available on the MUSYC 
website\footnote[6]{http://www.astro.yale.edu/MUSYC/}. We use four 
$UBVRIzJHK$ catalogs of the MUSYC fields i.e., the multiwavelength 
catalogs of the deepest subfields in E-HDFS ($201.1$ sq. arcmin), 
SDSS~1030 ($106.7$ sq. arcmin) and CW~1255 ($101.9$ sq. arcmin)
presented in \citet{Quadri2007} and the catalog of the full E-CDFS 
($969.6$ sq. arcmin; Taylor et al. 2009 in preparation). For each field, 
we select the p$BzK$, p$BzK$* and s$BzK$ galaxies to the 90\%\ completeness limit 
of our data. Surface densities of the MUSYC fields are given
in Table 2 (columns $3-6$). The surface densities derived from the four MUSYC 
fields is given in column $7$. Since GOODS-MUSIC is in E-CDFS, 
the surface densities for those two fields are not independent; see Table2,
columns $2-3$. 

Whereas the star-forming s$BzK$ galaxies only vary by up to 70\%\ from field to field,
the red p$BzK$ and p$BzK$* galaxies show significant field to field variations. 
The surface densities of both p$BzK$ and p$BzK$* galaxies around 7C~1756+6520 
are comparable to the MUSYC SDSS~1030 field, the denser control field as far 
as the red galaxies are concerned. We find an excess of p$BzK$ and p$BzK$*
galaxies by a factor of $2.4 \pm 1.2$ and $1.7 \pm 0.4$
relative to the average density derived from the four MUSYC fields. The density of 
s$BzK$ in the full 7C~1756+6520 field is, on the contrary consistent with the control fields. 
We find that $BzK$ densities in 7C~1751+6809 are all in good agreement with MUSYC 
and GOODS-MUSIC.

Overdensities of narrow-band emitters and EROs have been found out to $1.75$---$2$~Mpc in 
protoclusters around HzRGs at $2 < z < 3$ \citep{Kurk2004B, Venemans2007}.  \citet{Best2003} 
studied the radial distribution of EROs around powerful radio-loud AGN at lower redshifts 
($z \sim 1.6$) and found overdensities on scales of at least $1$~Mpc for four fields out of the 
six in their sample. We therefore also compute the density of sources found within $2$~Mpc 
($\sim 4\arcmin$) of the radio galaxies (Table 2, columns $9$ and $11$). The corresponding 
region is also outlined by dashed lines in Fig.~\ref{BzK2}. 
No excess of $BzK$ galaxies is found in the close surroundings of 7C~1751+6809. This region 
is actually under-dense for elliptical candidates by a factor of two compared to the full field. 
A concentration of galaxies is found in the $2$~Mpc surroundings of 7C~1756+6520 for p$BzK$,
p$BzK$* and s$BzK$ galaxies not only compared to the control fields - by a factor of 
$4.7 \pm 2.4$, $3.1 \pm 0.8$ and $2.0 \pm 0.6$
respectively - but also compared to the full 7C~1756+6520 field - by a factor of  
$1.7 \pm 0.9$, $1.9 \pm 0.5$ and $2.0 \pm 0.6$ respectively. We note that the full field 7C~1756-6520
is biased towards higher densities since it contains the excess of sources near the HzRG. 
If we compute the surface density of p$BzK$* in the rest of the field removing the $2$~Mpc surroundings
of the HzRG, the surface density reduces to $480 \pm 80$ sources per sq. degrees implying that the inner $2$~Mpc region is 
denser by a factor of $2.4 \pm 0.7$ compared to the rest of the field.

\begin{figure}[!t] 
\begin{center} 
\includegraphics[width=6.4cm,angle=90,bb=95 95 530 700]{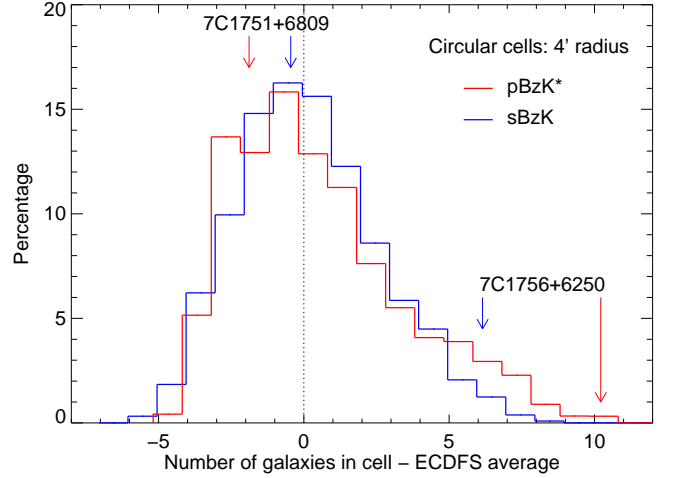}
\end{center}
\caption{Histogram of the number of p$BzK$* (red) and s$BzK$ (blue) galaxies in the MUSYC E-CDFS 
field in circular cells of $4\arcmin$ radius (equivalent to $2$~Mpc at z=1.48). We subtract from
our counts the number of galaxies expected in such a cell corresponding to the E-CDFS average 
density ($7.05$ for s$BzK$ and $5.18$ for p$BzK$*). Cells below (above) $0$ are therefore 
underdense (overdense) compared to the full
field. Arrows indicate the number of galaxies found within $2$~Mpc of our two HzRGs. We first note
that the tail of the red galaxies distribution falls further than the blue galaxies one, confirming that 
red galaxies are more clustered than blue ones. 7C~1751+6809 is found in a slightly underdense 
region for both blue and red $BzK$. The number of p$BzK$* for 7C~1756+6520 falls at the very end of the tail 
of the distribution with only $0.26$\% of the cells containing such a high number of red objects.}
\label{cell}
\end{figure}

In order to further quantify the probability to find an overdensity of p$BzK$* and s$BzK$ galaxies 
in a $2$~Mpc radius region, we now work out the counts-in-cells fluctuations of E-CDFS, the largest field of MUSYC. 
We measure the number of p$BzK$* and s$BzK$ galaxies (in the limits of completeness) 
found in $10000$ randomly placed circular cells of $4\arcmin$ radius (corresponding 
to $2$~Mpc radius at $z=1.48$) in the E-CDFS $969.6$ sq. arcmin field of view. Edges 
were avoided by forcing the cells centers to be at least $4\arcmin$ distant from the 
edges of the E-CDFS field. We chose a large number of cells (allowing some overlapping) 
to fully sample the counts fluctuations. We do not consider counts-in-cell of p$BzK$ 
in this analysis due to the very small number of these galaxies in E-CDFS ($18$).
Fig.~\ref{cell} shows the histogram of counts-in-cells.  s$BzK$ are more common than p$BzK$*.
In order to be able to directly compare those two populations, we subtract from our counts
the expected average density in E-CDFS scaled to the cell size. Counts are given in 
percentage of the total number of cells. We also mark with arrows the galaxies counts within 
$2$~Mpc of our two HzRGs (also corrected from the average E-CDFS density).
The histogram has a right-skewed distribution. 
We note that the tail of the distribution of red galaxies is longer that the one for blue galaxies
confirming that red galaxies are more clustered than blue ones \citep[see~][]{Daddi2000, Kong2006}.
Counts in 7C~1751+6908 are consistent, if anything slightly lower, than the average of E-CDFS. The 
counts of p$BzK$* and s$BzK$ galaxies around 7C~1756+6520 on the contrary fall way beyond 
the average density in E-CDFS, near the end of the tail of the distribution with only 
$0.26$\% of the cells having similar densities, confirming the result that the HzRG is found 
in an exceptionally overdense region.

\subsection{Number counts}

\begin{figure}[!t] 
\begin{center} 
\includegraphics[width=6.6cm,angle=90,bb=75 85 540 720]{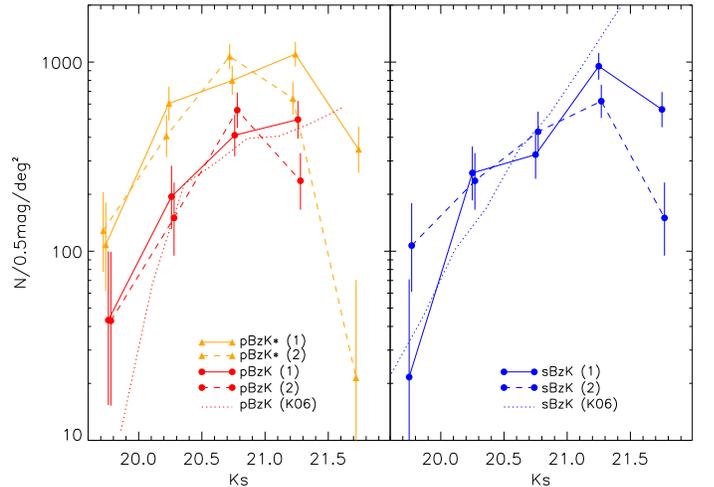}
\end{center}
\caption{$Ks$ number counts for p$BzK$ (red), p$BzK$* (orange) and s$BzK$ galaxies (blue)
compared with counts from Kong \etal (2006; K06).  The solid and dashed lines correspond
to 7C~1756+6520 (1) and 7C~1751+6809 (2), respectively.}
\label{Kcounts}
\end{figure}

\begin{figure*}[!t] 
\begin{tabular}{c c} 
\includegraphics[width=8.8cm,angle=0,bb=0 10 570 610]{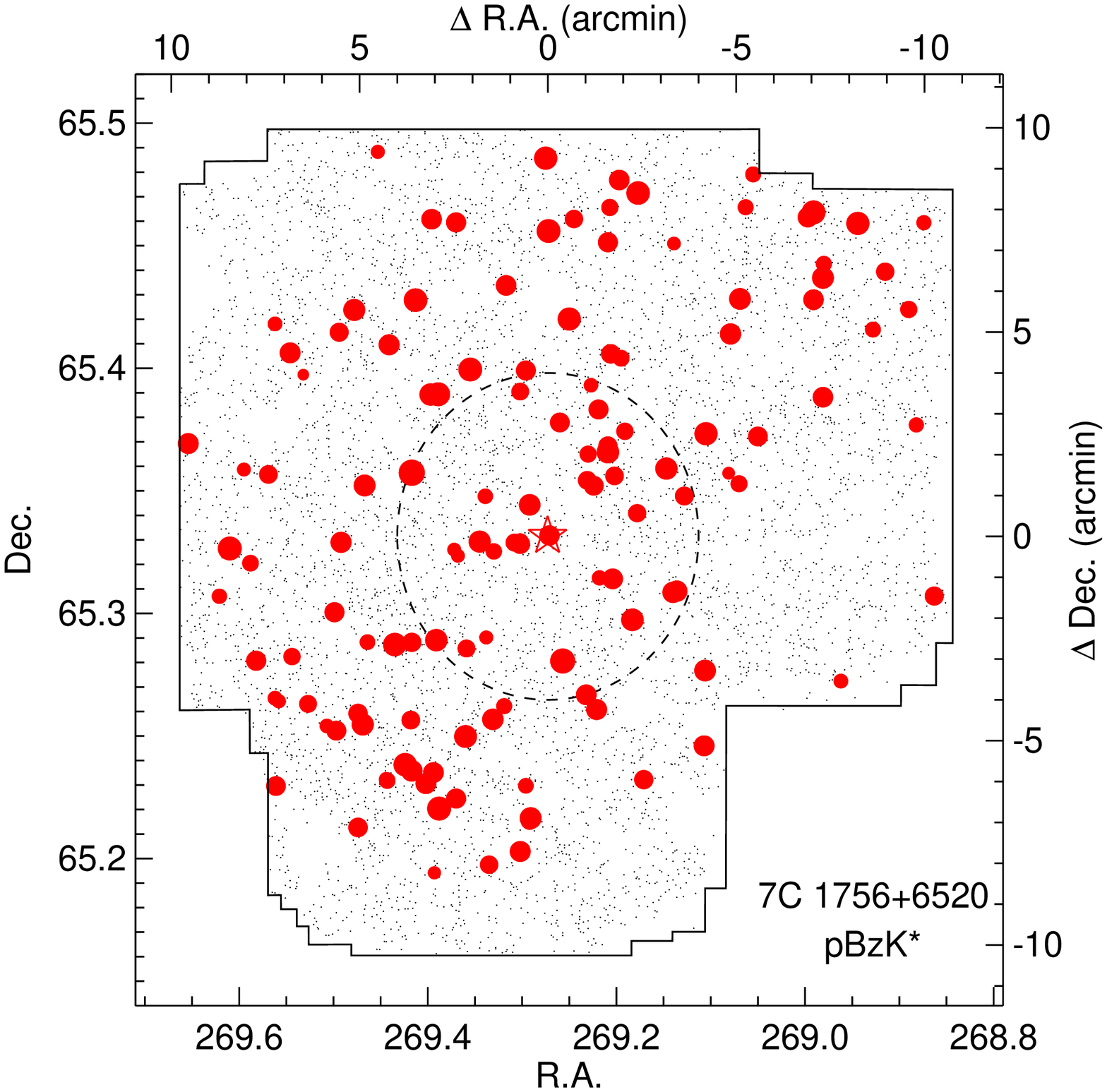} &
\includegraphics[width=8.8cm,angle=0,bb=0 10 570 610]{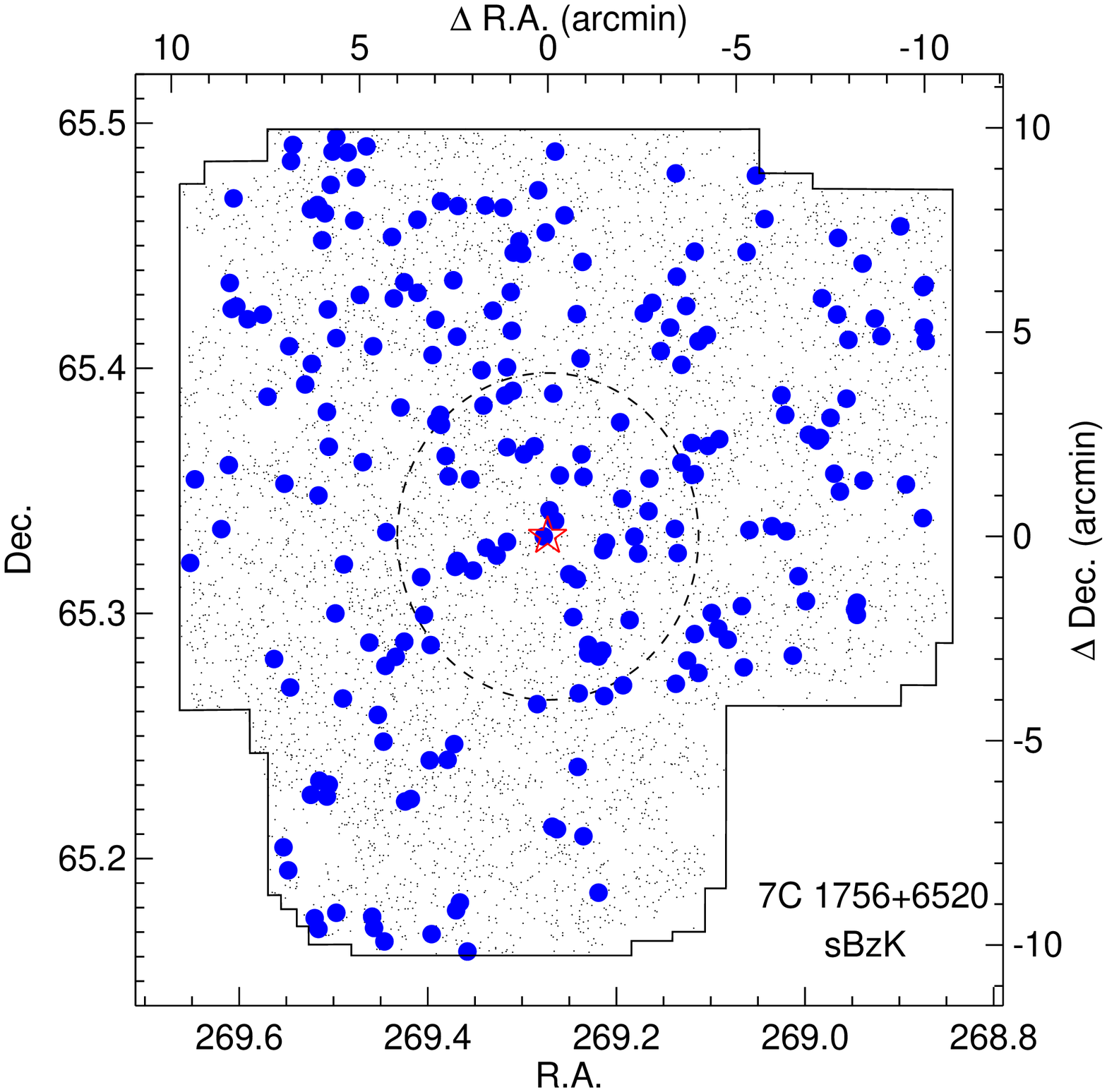}      \\
\includegraphics[width=8.8cm,angle=0,bb=0 10 570 610]{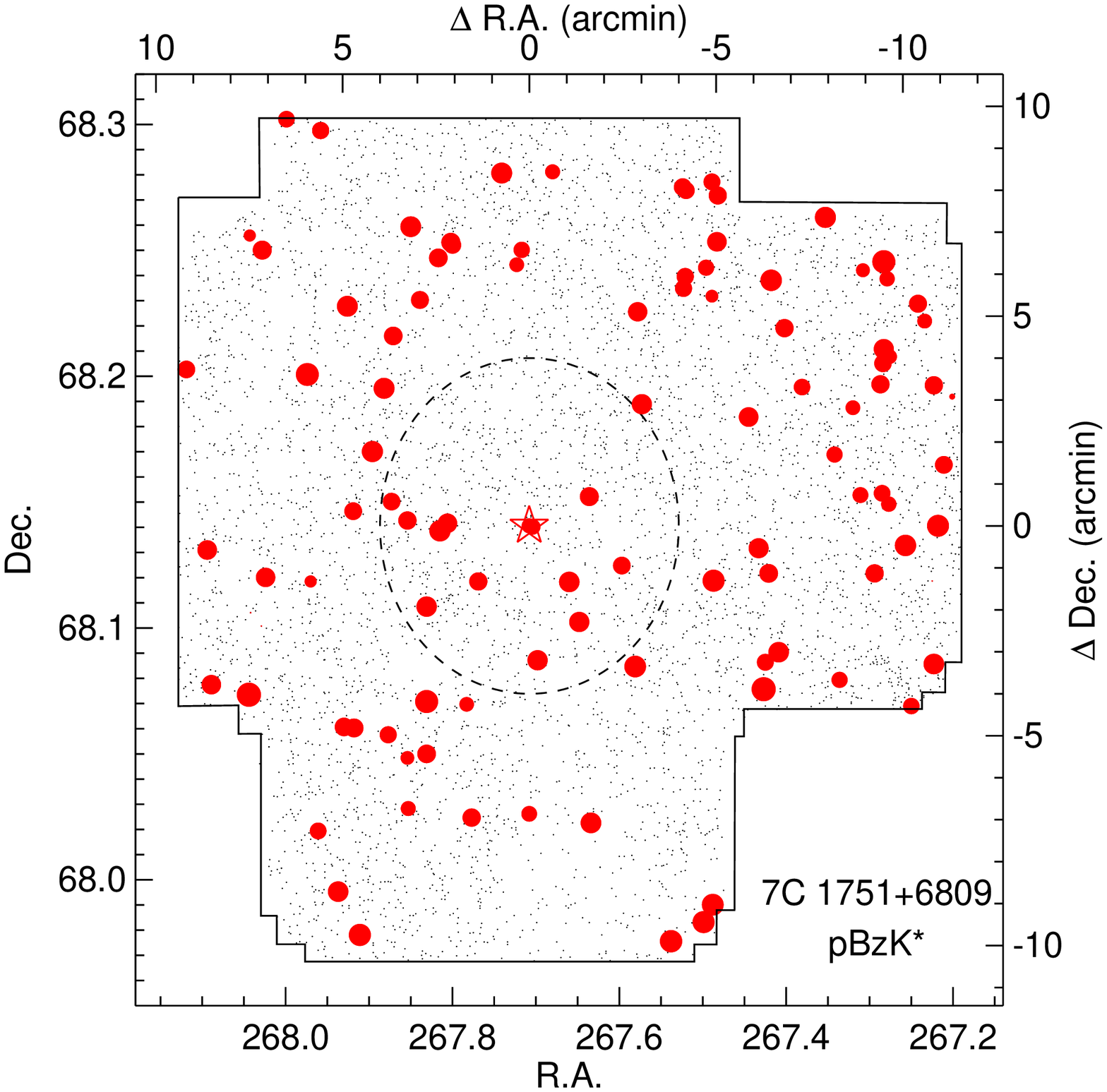} &
\includegraphics[width=8.8cm,angle=0,bb=0 10 570 610]{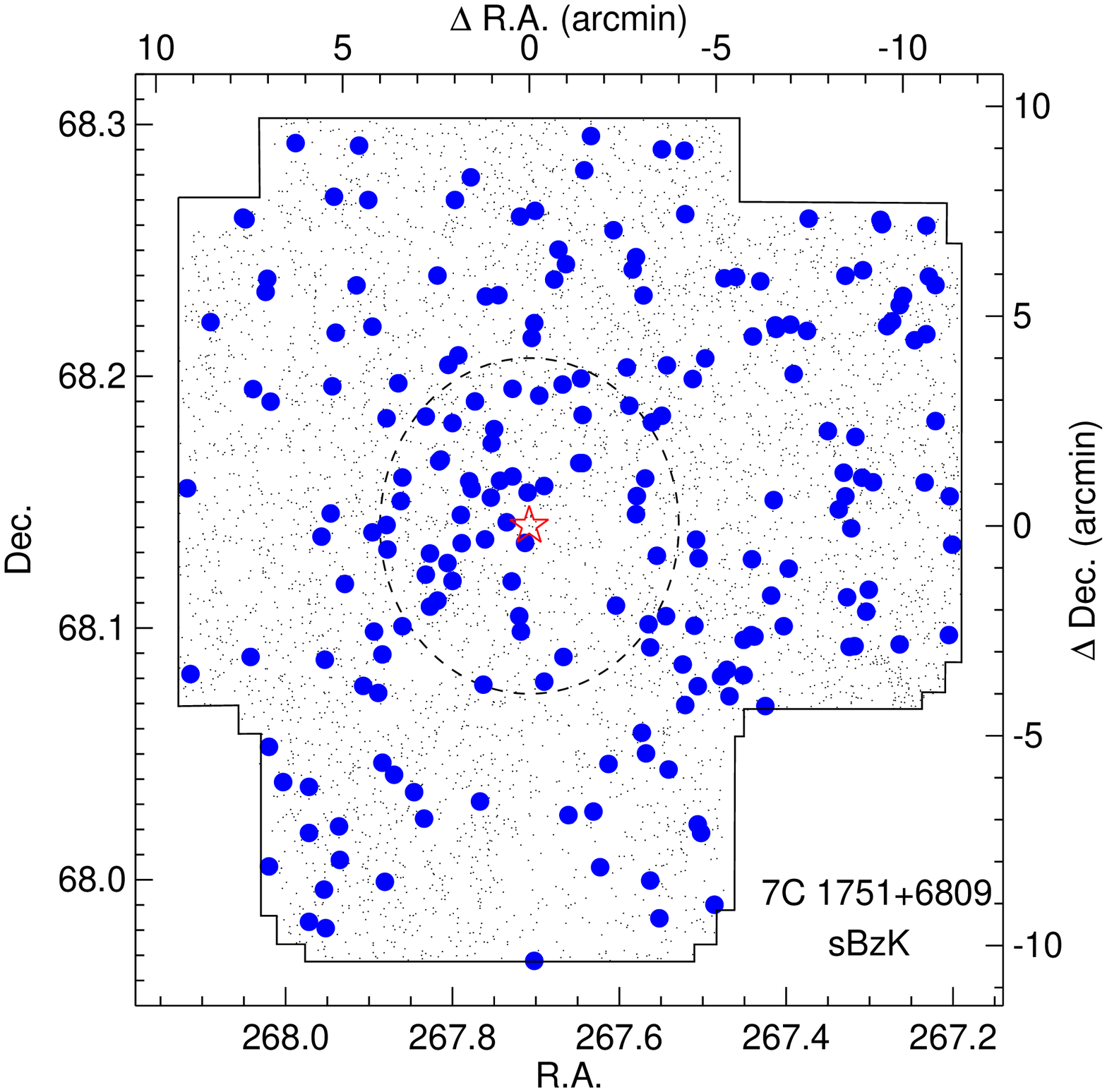} \\
\end{tabular}
\caption{Spatial distribution of the $BzK$-selected galaxies in the two fields: left panels, p$BzK$* galaxies
in red with the size of the symbol scaled according to the $Ks$-magnitude; right panels, s$BzK$ galaxies in blue. 
The radio galaxies are marked by the red stars. The area simultaneously covered by the $BzK$ bands is 
outlined and all sources detected in all three bands ($3\sigma$) are indicated by black points. The coordinates 
are also given relative to the radio galaxy in arcmin (right and top axes). The $2$~Mpc radius region around the HzRG
is marked by the dashed line.}
\label{BzK2}
\end{figure*}

We derive the $Ks$-band number counts in $0.5$ mag bins for p$BzK$, p$BzK$* and s$BzK$ galaxies in
our two fields (Fig.~\ref{Kcounts}). We adopt Poissonian
errors for the counts and use the \citet{Gehrels1986} small numbers approximation for Poisson 
distributions. We overplot the findings of Kong et al. (2006; K06 hereafter) as a dotted line for comparison 
\citep[see also ][]{Lane2007, Imai2008, Hartley2008}. The number counts become incomplete 
at $Ks > 21$ when we start reaching the completeness limit of our $z$-band. 

The number counts of p$BzK$ and s$BzK$ galaxies are in good agreement with 
\citet{Kong2006}. We find the number of s$BzK$ galaxies increasing steeply with 
decreasing magnitude, with the slope for the 7C~1756+6520 field similar to K06. 
7C~1751+6809, however, shows a small excess ($2\sigma$) of s$BzK$ galaxies 
at bright $Ks$ magnitudes ($19.5 < Ks < 20.5$). The slope of the number counts 
for p$BzK$ galaxies is similar for both of our radio galaxy fields and \citet{Kong2006}. 
A small excess of $Ks$-bright p$BzK$ galaxies is suggested in both fields ($1.5\sigma$). 
Such excesses of $Ks$-bright galaxies have also been noticed around other HzRGs 
\citep{Kodama2007}. If these $Ks$-bright sources were associated with the HzRGs, 
they would be very massive ($M > 10^{11} M_{\sun}$) and would represent the massive, 
evolved galaxy population of a young galaxy cluster around the HzRG. However, the number 
of sources considered here is too small to reach any firm conclusion since we detect only 
five p$BzK$ galaxies with $19.5< Ks < 20.5$.  Previous work \citep[K06; ][]{Lane2007, Hartley2008} 
has shown that the p$BzK$ number counts show a turn over at $K > 21$ with the counts 
slope flattening at fainter sources. We also strongly suspect this flattening in our counts 
although our $BzK$ selection rapidly becomes incomplete at $Ks \geq 21$. A possible 
explanation for this turn-over is that p$BzK$ galaxies are selected in a small redshift range 
and consist of very massive, passively evolving galaxies. Due to downsizing, their number 
decrease at lower luminosities \citep{Hartley2008}. The slope of the counts and the range of 
$Ks$ magnitudes sampled by the p$BzK$* galaxies are also consistent with the p$BzK$ 
galaxies, suggesting that our extended selection criteria is also likely to be selecting 
galaxies in the same redshift range.


\section{Properties of candidate massive cluster members}

\subsection{Spatial distribution}

Fig.~\ref{BzK2} displays the spatial distribution of the candidates around the 
radio galaxies. We also plot all the sources detected ($3\sigma$) in $B$, $z$ 
and $Ks$ (black dots) in order to visualize the zones of the field affected by 
our cross-talk flag (see Fig.~\ref{weightmap}) and very bright stars. In both 
fields, the $BzK$ source distribution is clearly inhomogenous. In the 
7C~1756+6520 field, the passive candidates (Fig.~\ref{BzK2}; top left panel) 
are more numerous in the surroundings of the radio galaxy than in the rest 
of the field. This was also seen in \S3.3 and Table 2. Eight p$BzK$* galaxies 
are roughly aligned with the HzRG in a small E/W structure $\sim 2.5'$ in length. 
Two other excesses are also observed in the field, one $\sim 3\arcmin$ SE of the 
HzRG and another one at $\sim 6.5\arcmin$ NW of the HzRG. These excesses 
appear to be aligned with a global overdensity of p$BzK$* galaxies along a large 
structure in a NW-SE direction. In contrast, no clear excess of p$BzK$* galaxies 
is detected in the 7C~1751+6809 field (Fig.~\ref{BzK2}; bottom left panel) even 
though they show a relatively non-uniform distribution. We note that one p$BzK$* 
galaxy is found near the line of sight to both HzRGs (at $2.4\arcsec$ for 7C~1756+6520
and $6.3\arcsec$ for 7C~1751+6809), suggesting that both HzRGs may have close 
companions.

The s$BzK$ galaxies also have an inhomogenous distribution though it is less well 
defined than the p$BzK$* galaxies. In particular, 7C~1756+6520 has nine s$BzK$ 
galaxies along the same elongated structure near the HzRG. One s$BzK$ galaxy 
is also found near the line of sight of the HzRG ($5.5\arcsec$, Fig.~\ref{Panel}). The 
s$BzK$ galaxies in the 7C~1751+6809 field show an excess near the center of the 
field with an elongation in the direction NW-SE with no obvious correlation with the 
p$BzK$* spatial distribution. This overdensity was not seen with the density counts 
in Table 2, most probably due to the fact that the HzRG is at the ``edge'' of the elongated 
structure of s$BzK$ galaxies. If associated with the HzRG, this structure would 
have a $\sim4$~Mpc extent. 

\begin{figure}
\begin{center} 
\includegraphics[width=9cm,angle=0,bb=10 15 560 570]{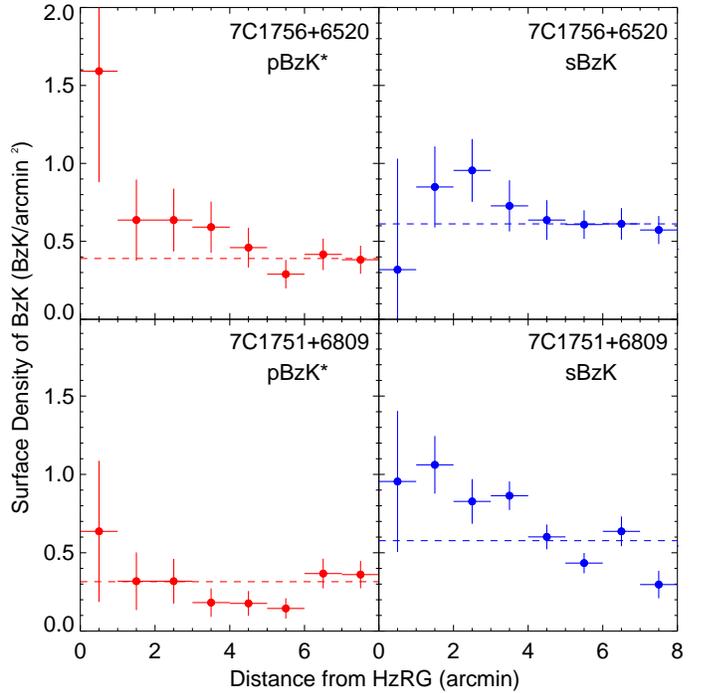} 
\end{center}
\caption{Radial density profile of $BzK$ selected galaxies around 7C1756+6520 (upper panels) and
7C~1751+6809 (lower panels) for p$BzK$* galaxies (left panels) and s$BzK$ galaxies (right panels). 
The full field density is shown by the horizontal dashed lines. The profiles and surface densities 
were derived from the entire sample of candidates. The values obtained are therefore
higher than in Table~1 where the study was restricted to the completeness limit. The error bars 
indicate the $1\sigma$ errors on the counts assuming Poissonian errors.}
\label{radius}
\end{figure}

Fig.~\ref{radius} presents the radial distribution of p$BzK$* and s$BzK$ galaxies around
both HzRGs i.e., the number of candidates found per radius bin (binsize$=1\arcmin$) divided by
the corresponding ring area. The large error bars are due to the small number of sources used
to derive the radial profiles, e.g., only $5$ p$BzK$* and $1$ s$BzK$ galaxies are found
within $1\arcmin$ of 7C~1756+6520. The distribution of candidates near 
7C~1756+6520 forms an elongated structure, not centered around the HzRG.
The radial profile in Fig.~\ref{radius} is therefore a lower limit to the true concentration of $BzK$
around the HzRG as it does not fully reflect the complex spatial distribution of the sources.
However, we note that a clear peak of p$BzK$* galaxies is seen near the HzRG. The p$BzK$* density 
decreases with radius and asymptotes to the full field density (red dashed line)
at $\sim5\arcmin$ from the radio galaxy. Variations are also observed in the profile of the s$BzK$ galaxies
with a deficit of sources near the HzRG ($<1\arcmin$) and a ``bump'' in the profile 
between $2\arcmin$ and $4\arcmin$, suggestive of some segregation in the properties
of the galaxies in the large scale structure. We note however that the significance of those 
variations is less that $1\sigma$. As seen previously, no significant variation of the p$BzK$* 
density is seen around 7C~1751+6809 but a small overdensity of s$BzK$ is observed 
within $5\arcmin$ of  the HzRG ($1\sigma$ significant though).


\subsection{Color-magnitude diagram}

\begin{figure*}
\begin{tabular}{c c}
\includegraphics[width=8.5cm,angle=0,bb=10 25 520 580]{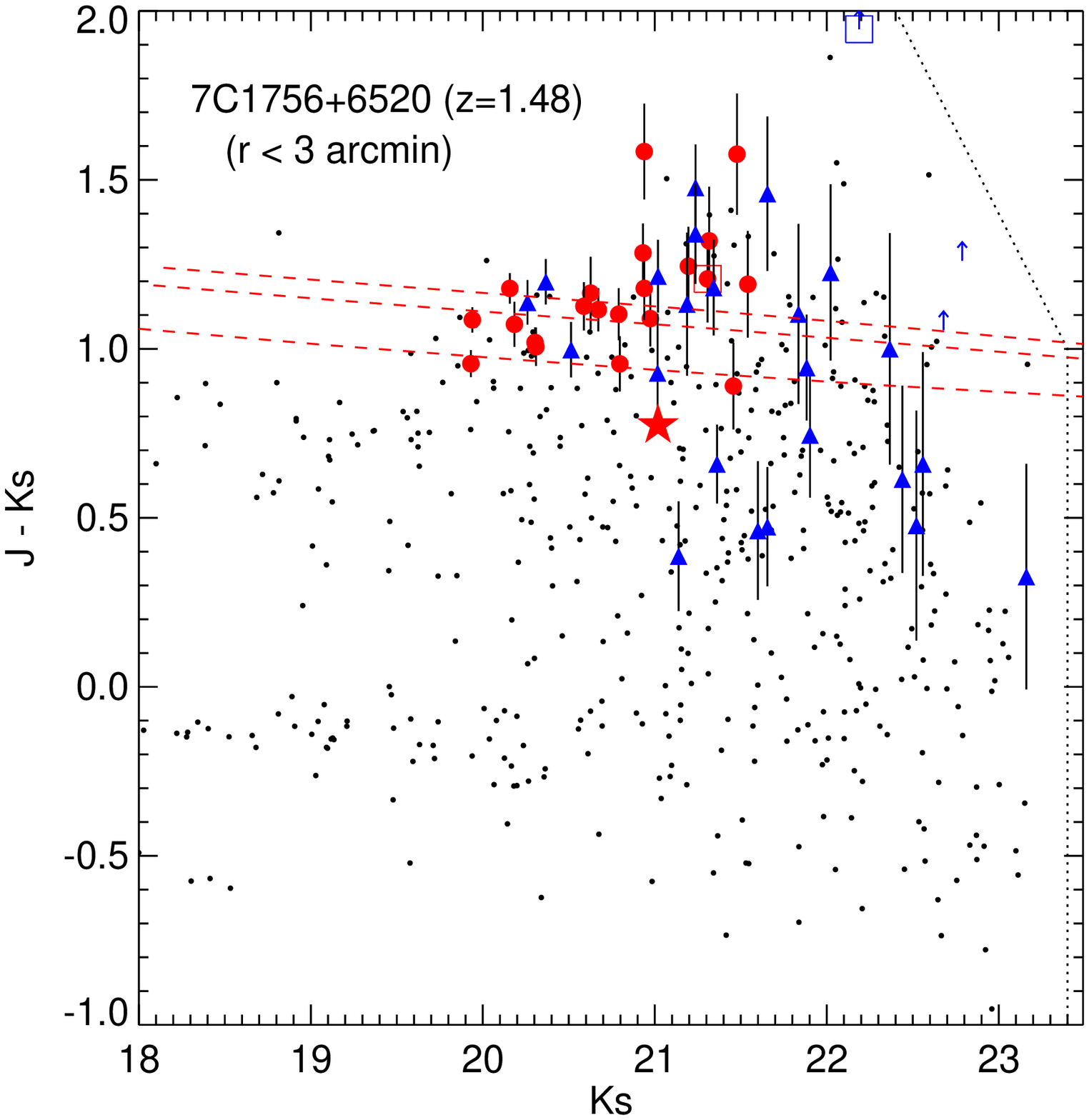} &
\includegraphics[width=8.5cm,angle=0,bb=10 25 520 580]{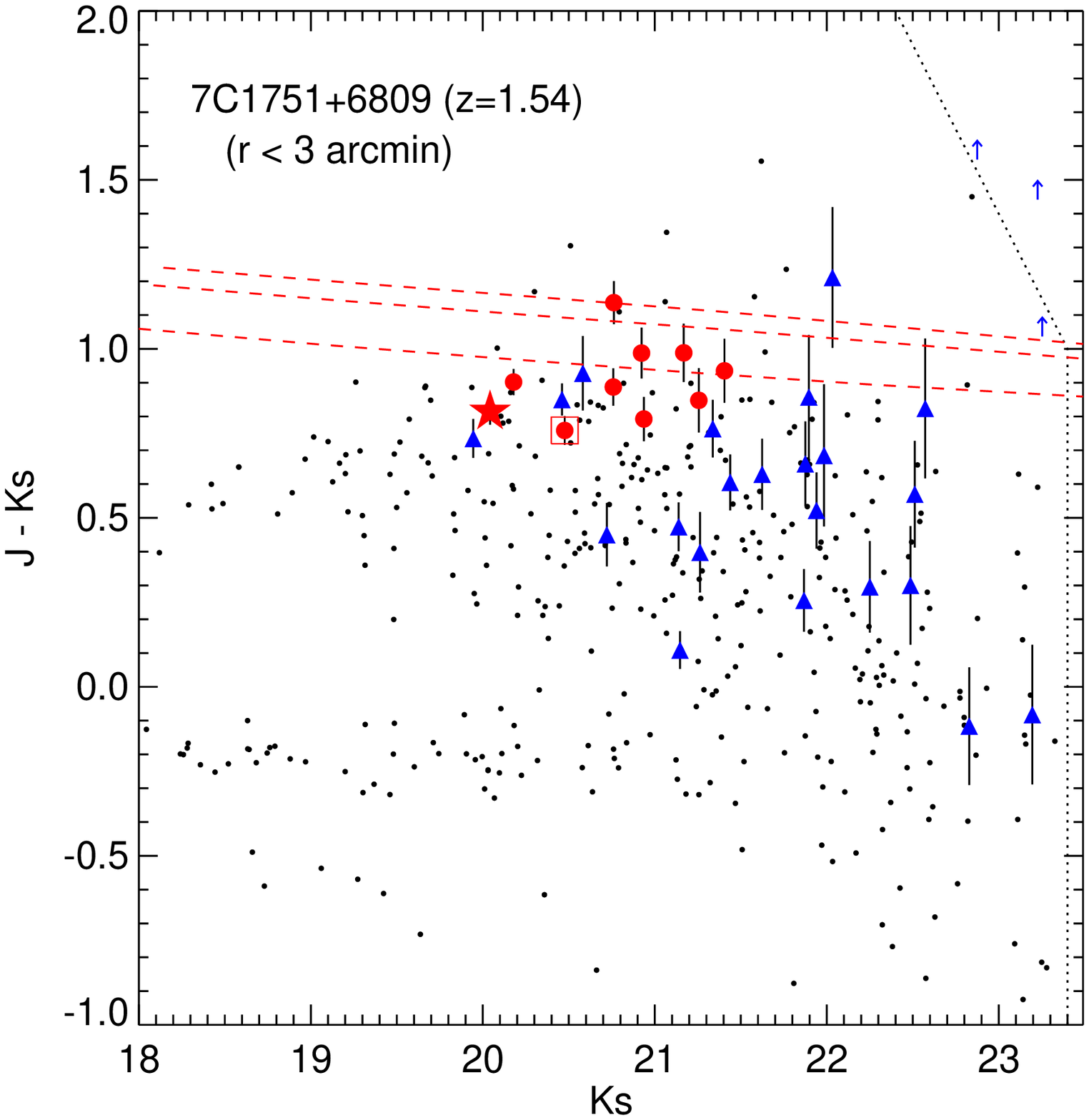} \\
\end{tabular}
\caption{Color-magnitude diagram ($J-Ks$ vs $Ks$) of the regions surrounding the HzRGs (within $3'$). 
The p$BzK$* and s$BzK$ galaxies are plotted as red circles and blue triangles respectively. s$BzK$ galaxies not detected
in $J$ are shown as blue arrows. Also plotted are all sources in this same region detected in the three 
$BzK$ bands (black dots). The radio galaxies are shown as red stars. The dotted lines represent the 
$3\sigma$ detection limits in $J$ and $Ks$ reported in \S2.2.3. The two p$BzK$ galaxies found near the line of 
sight of both HzRGs and the s$BzK$ galaxy found near 7C~1756+6520 are indicated by the red and blue squares, 
respectively. The dashed lines indicate the expected locations of the red sequence at $z = 1.5$ corresponding to the predicted color
of a passively evolving stellar population with $z_f = 3, 4$ 
and $5$ (from lower to upper curve; see text for details).}
\label{CMD}
\end{figure*}

Color-magnitude diagrams (CMDs) are an efficient method to study the 
formation and evolution of galaxies. At $z < 1$, galaxy cluster cores are 
dominated by massive, passively-evolving elliptical galaxies that trace 
a clear red sequence on the color-magnitude diagram. In the last decade, 
studies have shown that this red sequence of early-type galaxies is also found in 
galaxy clusters out to $z\sim1.5$ \citep[e.g.~][]{Mei2006, Stanford2006, 
Tanaka2007, Lidman2008}. Recent work at even higher redshifts have 
studied the evolved galaxy population in $z\sim2$ galaxy clusters and 
conclude that the red sequence may appear between $z = 3$ and $2$ 
\citep{Kodama2007, Zirm2008}.  We have investigated the CMD of the 
sources in the region surrounding the HzRGs. Their CMDs are shown in 
Fig.~\ref{CMD}. Sources within $3'$ of the HzRGs and with a $3\sigma$ 
detection in all $BzK$ bands are plotted as black dots. The size of the 
studied region was chosen as a compromise between selecting sources 
close to the HzRG and including the majority of the candidates in the 
apparent central overdensity. We note that the region of the CMD at faint 
$Ks$ magnitude starts to be empty well before the magnitude limit of our 
$Ks$-band data due to the non detection of faint $Ks$ sources in the 
optical bands. p$BzK$* and s$BzK$ galaxies within $3'$ ($\sim 
1.5$~Mpc at $z=1.5$) are plotted as red points and blue triangles, 
respectively. All p$BzK$* galaxies found near the HzRGs have a 
$> 3\sigma$ detection in the $J$-band; s$BzK$ galaxies with lower limits 
in $J$ are marked as blue arrows. The two p$BzK$* galaxies found near 
both HzRGs and the s$BzK$ galaxy found near 7C~1756+6520 are marked 
as squares. We overplot models of the expected location of the red sequence 
at $z=1.5$, \ie the predicted $J-K$ color of a passively evolving galaxy 
with different formation redshifts ($z_f$ = 3, 4, 5; provided by T.Kodama). 
The models reproduce the red sequence of passively evolving galaxies in 
the Coma cluster at $z=0$ and include a metallicity-magnitude dependance 
which causes the red sequence slope \citep{Kodama1998}. 

The p$BzK$* galaxies in the inner $3\arcmin$ region around 7C~1756+6520 
have colors consistent with passively evolving galaxies with $z_f \geq 2$ in 
contrast to the p$BzK$* galaxies around 7C~1751+6809 which have bluer 
$J-Ks$ colors. Some elliptical candidates have slightly redder colors ($J-Ks > 1.2$) 
and may be background objects since the $BzK$ criteria is designed to select objects 
at $1.4 < z < 2.5$. Two of the p$BzK$* and three of the s$BzK$ galaxies have 
$(J-K)_{\rm Vega} > 2.3$ and would be classified as DRGs, \ie they are likely to be either passive 
elliptical or dusty star-forming galaxies at $z > 2$.

Recent observations of some high redshift galaxy clusters have shown a 
deficit of red galaxies at the faint end of the red sequence compared to local 
clusters \citep{Kajisawa2000, DeLucia2007, Tanaka2005, Tanaka2007}. It has 
been suggested that the red sequence appears at bright magnitudes and 
progressively extends to fainter magnitudes with time. \citet{Tanaka2007} studied 
a possible large-scale structure around a galaxy cluster at $z = 1.24$ and found 
that a deficit of faint red galaxies is noted in the clumps surrounding the central 
cluster but not in the CMD of the cluster itself, suggesting that the build-up of the 
red sequence is dependent on environment, progressing more rapidly in higher 
density environments. Considering the potential cluster around 7C~1756+6520, 
we note a clear deficit of $Ks$-faint p$BzK$* galaxies. No p$BzK$* galaxy is found 
with $Ks > 21.5$ near the HzRG. At these faint $Ks$ magnitudes, we surely reach 
the combined incompleteness of our $z$ and $Ks$ bands data. But, as described 
in \S3.2, we are more than $60$\% complete at our magnitude limits. For example, 
$29$ p$BzK$* galaxies with $Ks > 21.5$ are found in the full field and s$BzK$ 
galaxies are found with $Ks > 21.5$ within $3'$ of the HzRG. We therefore conclude
that the truncation at faint magnitudes is real. This would imply that this is another example of 
downsizing \citep{Cowie1996}; i.e., the more massive cluster members stopped their star-formation
earlier than the less massive cluster members. A similar study of the CMD of red galaxies in the 
field of the X-ray galaxy cluster XMMUJ2235.3-2557 at $z=1.39$ is presented in \citet{Lidman2008}.
They do not observe evidence of a truncation of the red sequence at fainter magnitudes, suggesting that
they are looking at a richer or more evolved system.
The scatter of the p$BzK$* galaxies relative to the red sequence model at $z_f = 5$ ($z_f = 4$) is 
$0.089\pm0.067$ ($0.095\pm0.061$) magnitudes for non-DRG galaxies. This scatter is large and
most probably inflated by non-cluster members. Studies of the intrinsic scatter of the red sequence 
in galaxy clusters at $1.2 < z < 1.5$ have however shown that the scatter in $J-K$ can be up to $\sim0.06$ 
\citep{Lidman2004, Lidman2008}. 

We stress that the p$BzK$* galaxies selected in this work are only candidate cluster members and that spectroscopic 
follow-up will be necessary to confirm their physical association to the HzRGs.

\section{AGN candidates}

Recent studies suggest that AGN companions are often found around radio galaxies. 
\citet{Croft2005}  spectroscopically confirmed three QSOs in the surroundings of 
PKS~1138-262 at $z=2.16$ and suggested that the QSOs were triggered by the 
protocluster formation \citep[see also~][]{Pentericci2000}.  \citet{Venemans2007} 
also detected QSOs near radio-galaxies at $z > 3$.  Recently, \citet{Galametz2009} 
studied the AGN population in a large sample of galaxy clusters at $z < 1.5$ and 
found an excess of AGN within $0.5$~Mpc of the cluster centers, with the number 
of AGN in clusters increasing with redshift \citep[see also~][]{Eastman2007}. Powerful 
AGN provide an alternative way to look for relatively massive host galaxies in a 
complementary technique to the near-IR color selection.

\begin{figure}[!t] 
\includegraphics[width=8.5cm,angle=0]{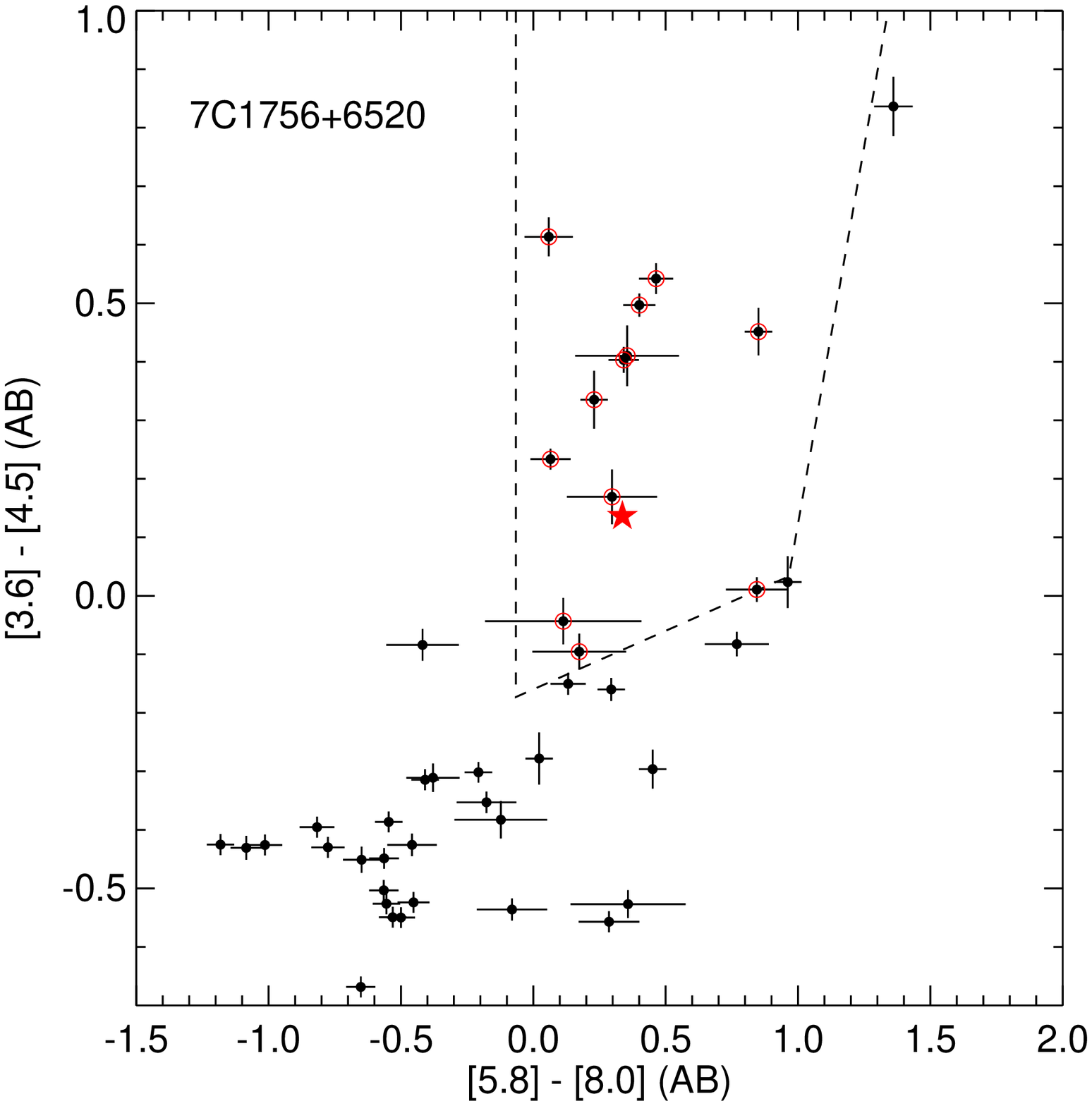} 
\includegraphics[width=8.5cm,angle=0]{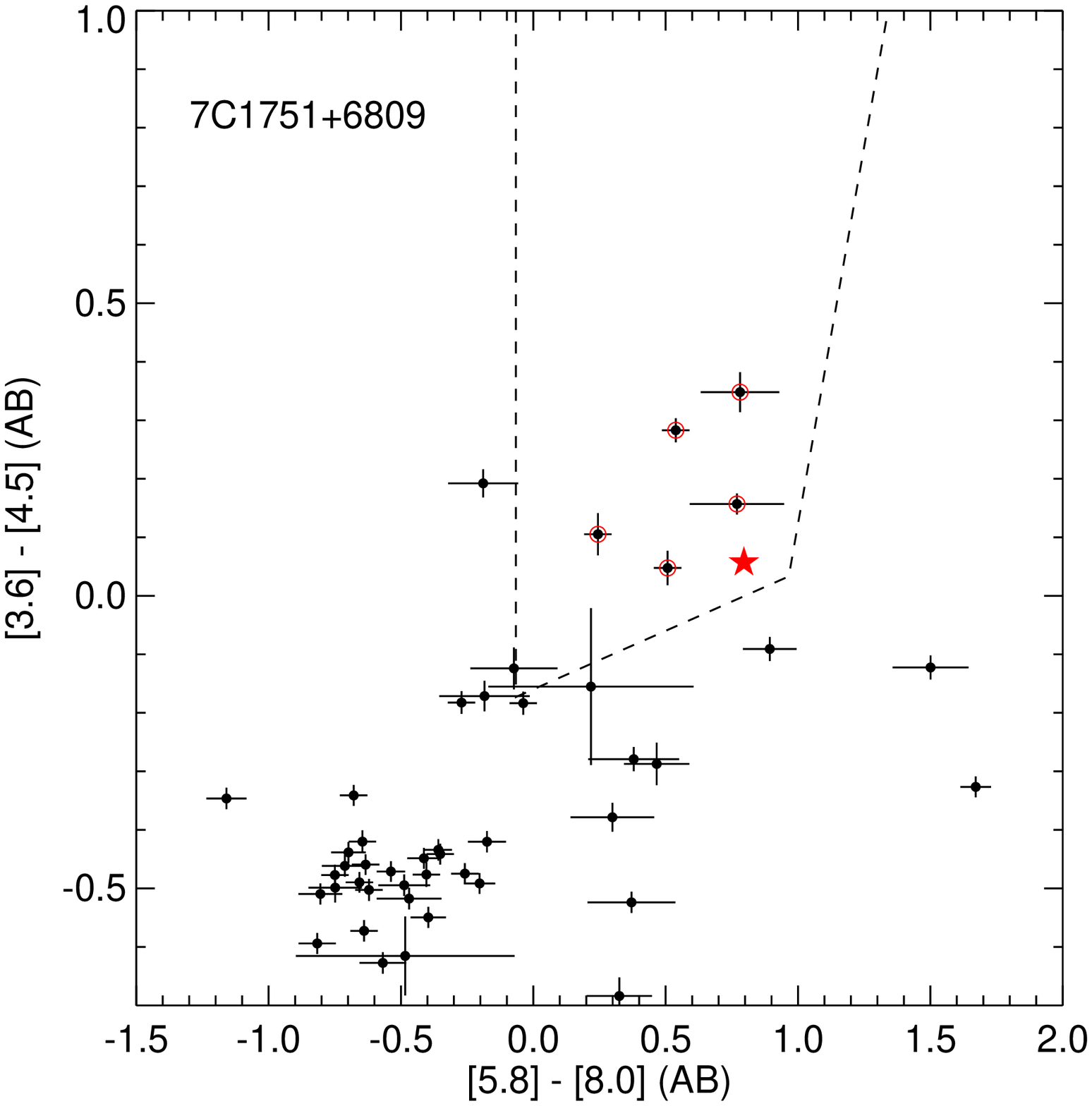} 
\caption{Mid-IR color-color diagram for 7C~1756+6520 (top) and 7C~1751+6809 (bottom).
All sources with a $5\sigma$ detection in all four IRAC bands are plotted. We overplot the \citet{Stern2005} wedge
for AGN selection. Sources circled in red are identified as AGN candidates by this criterion. HzRGs are indicated
by the red stars and are both found in the selection wedge, as expected.}
\label{AGN}
\end{figure}

\citet{Stern2005} presents a robust technique for identifying active galaxies from 
mid-infrared color criteria \citep[see also~][]{Lacy2004}.  While the continuum 
emission of stellar populations peaks at approximately $1.6\mu$m, the continuum 
of AGN is dominated by a power law throughout the mid-infrared. \citet{Stern2005} 
adopt the following (Vega system) criteria\footnote[7]{We use the following conversions 
between the Vega and AB photometric systems: $[3.6]_{\rm AB} = [3.6]_{\rm Vega} + 2.792$, 
$[4.5]_{\rm AB} = [4.5]_{\rm Vega} + 3.265$, $[5.8]_{\rm AB} = [5.8]_{\rm Vega} + 3.733$ and
$[8.0]_{\rm AB} = [8.0]_{\rm Vega} + 4.399$.} to isolate AGN from other sources: 
$([5.8]-[8.0]) > 0.6 \cap ([3.6]-[4.5]) > 0.2 \times ([5.8]-[8.0])+0.18 \cap ([3.6]-[4.5]) > 2.5 \times ([5.8]-[8.0])-3.5$
Since the criterion is designed to identify power-law spectra, they do not preferentially
select AGN in any specific redshift range.
We apply this selection criteria to all sources with a $5\sigma$ detection in all four IRAC bands. 
The coordinates of the selected AGN for 7C~1756+6520 ($12$ candidates) and 7C~1751+6809
($5$ candidates) are given in Table~\ref{AGN7C1756} and Table~\ref{AGN7C1751}, respectively. 
Fig.~\ref{AGN} shows their distributions in the [3.6]-[4.5] vs [5.8]-[8.0] color-color diagram.
We note that although neither HzRG is detected at a $5\sigma$ level in the $5.8\mu$m-band,
their IRAC magnitudes and position in the IRAC color-color diagram are presented in the tables 
and in Fig.~\ref{AGN}. Both are undeniably classified as AGN by the \citet{Stern2005} criterion.

As a comparison field, we use the IRAC Shallow Survey \citep[ISS;][]{Eisenhardt2004}, which 
includes four IRAC bands and covers $8$ square degrees in the Bo\"otes field with at least $90$s 
exposure time per position. $2262$ sources in ISS are found in the \citet{Stern2005} AGN selection 
wedge, where we require a $5\sigma$ detections in all four IRAC bands \citep{Galametz2009}.
The $5.8\mu$m band is the least sensitive with a $5\sigma$ limiting depth of $15.9$ (Vega, in an
aperture-corrected $3\arcsec$ diameter aperture; equivalent to $51 \mu$Jy). Whereas we would 
expect three to four AGN candidates in the HzRG fields, we find $8$ AGN candidates near 
7C~1756+6520 and four near 7C~1751+6809  at the depth of ISS. We therefore observe an 
overdensity of AGN candidates in the field of 7C~1756+6520 by a factor of two compared to 
the 7C~1751+6809 and ISS fields. One AGN candidate is found only $5\arcsec$ offset from 
7C~1756+6520 (Fig.~\ref{Panel}, \S6) and two additional candidates are found within $1.5\arcmin$
of the HzRG. However, the $12$ AGN candidates do not show any particular spatial distribution 
as was seen for both the p$BzK$* and s$BzK$ galaxies (see \S4.2). No AGN candidate
is found within $1.5\arcmin$ of 7C~1751+6809.


\section{Candidate close companions to 7C~1756+6520}

\begin{figure*}[!t] 
\begin{center} 
\includegraphics[width=12cm,angle=90,bb=100 200 520 620]{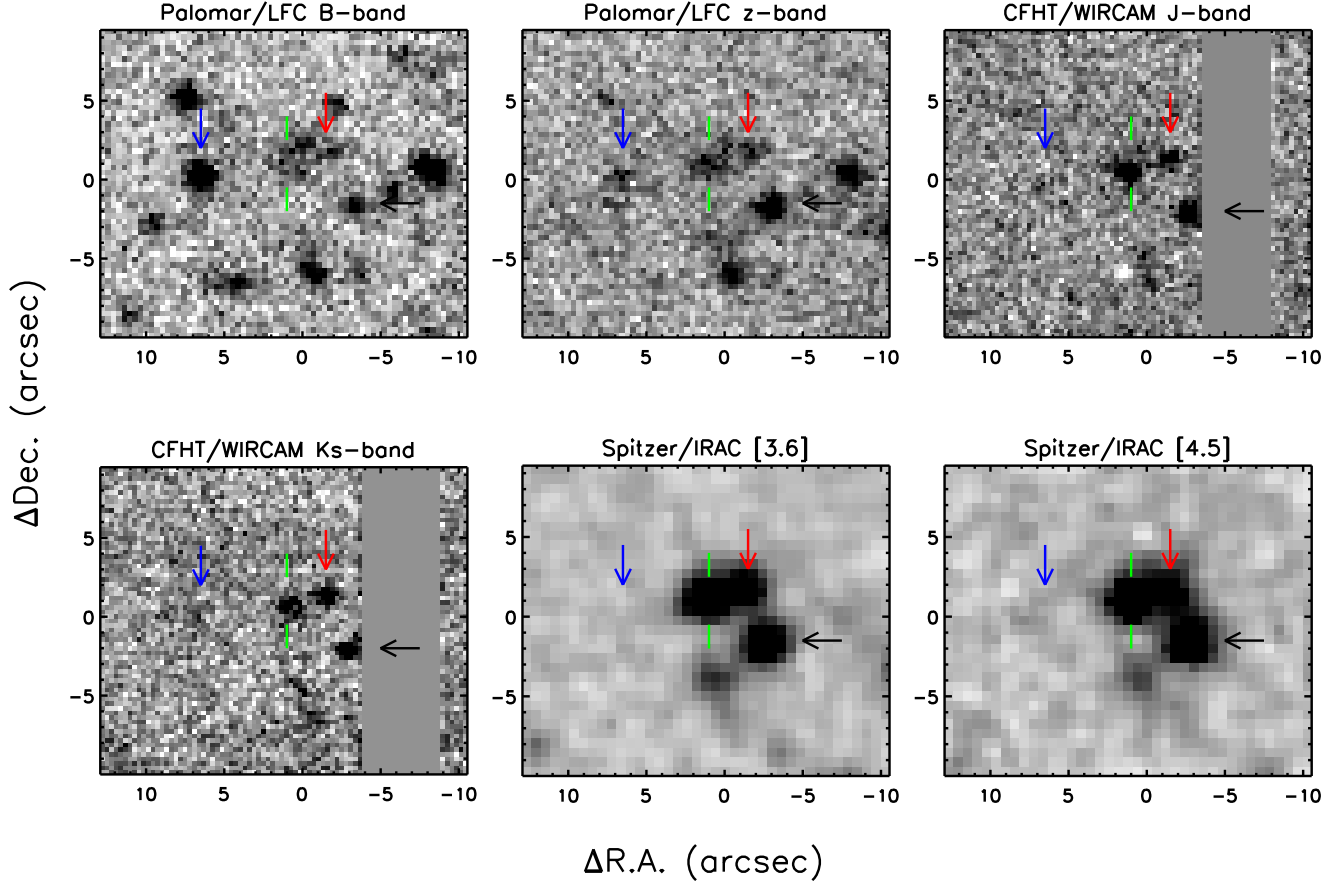}
\end{center}
\caption{7C~1756+6520 and its immediate surroundings in our Palomar/LFC $B$ and $z$-bands,
CFHT/WIRCAM $J$ and $Ks$-bands, and Spitzer/IRAC $3.6\, \mu$m and $4.5\, \mu$m images. North is up and East is to the left.
7C~1756+6520 is indicated by the green marks. The s$BzK$ galaxy, p$BzK$ galaxy and AGN candidate found near the 
HzRG are marked by the blue, red and black arrows, respectively.}
\label{Panel}
\end{figure*}

An elliptical, a star-forming and an AGN candidate are found near the line of sight
to 7C~1756+6520 (within $6\arcsec$), suggestive of several close
companions. Fig.~\ref{Panel} shows the immediate surroundings of 7C~1756+6520 in 
our imaging bands; arrows indicate the p$BzK$* galaxies, s$BzK$ galaxies and the AGN candidate. 
Using the density of $BzK$ galaxies found in the full field, we find that the 
probability of finding a p$BzK$* galaxy within $6\arcsec$ of the HzRG is $\sim0.44$\%, 
and the probability of finding a s$BzK$ galaxy is $\sim0.54$\%. At the depth of our IRAC data, the probability of 
finding an AGN candidate in the same area is $\sim0.57$\%. The probability of finding the 
three candidates in this small area around the HzRG is therefore extremely small,
strongly suggesting that these candidates are associated with the HzRG and
form a very unique and diverse system of bound galaxies.

\section{Conclusions}

We study of the surroundings of two radio galaxies at $z \sim 1.5$ using deep
multiwavelength imaging. We select candidate cluster members using color 
selection techniques designed to select galaxies at the redshift of the targeted 
HzRG. This technique has been proven to identify clusters and proto-clusters 
at high redshift \citep[$z > 2$;~][]{Kajisawa2006,Tanaka2007}. 
An excess of candidate passive elliptical candidates is found in the field of one of our 
two targets, 7C~1756+6520 by a factor of $3.1 \pm 0.8$ compared to control fields. A 
study of the counts-in-cells fluctuations in our larger control field shows that the probability 
to find such an overdensity in the field is very low ($0.26$\%). These results may 
be compared to previous studies that have been made at similar or higher redshifts. 
The \citet{Best2003} study of the environments of six radio-loud AGN at $z \sim 1.6$ 
finds an excess by a factor of $1.5$ to $4$ of EROs within radial distances of $\sim1$ 
Mpc of the AGN. Similarly, \citet{Kodama2007} select passive evolving and star 
forming cluster member candidates in the surroundings of four HzRGs with 
$2 < z < 3$ applying color cuts in $JHKs$ and found excesses by a factor of two 
to three compared to the field. Clusters were already suspected around those 
HzRGs in previous studies that were concentrated on overdensities of narrow-band 
(H$\alpha$, Ly$\alpha$) emitters also by a factor of two to five larger than in the field 
\citep{Kurk2004A, Venemans2005, Venemans2007}. Looking at narrow-band 
emitters has been very efficient at finding overdensities around HzRGs. However, such 
clusters members, dominated by young stellar populations, are likely not the most 
massive members of the galaxy clusters. Indeed, recent studies show that 
Ly$\alpha$ emitters have rather small stellar masses. \citet{Finkelstein2007} found 
masses ranging from $2 \times 10^7$ to $2 \times 10^9 M_{\sun}$ for a sample of $98$ 
Ly$\alpha$ emitters at $z\sim4.5$. Similar masses were deduced from Ly$\alpha$ emitters in 
\citet{Gawiser2007} who found stellar masses of $10^9 M_{\sun}$ for lower redshift 
objects ($z\sim3.1$). Looking at the properties of Ly$\alpha$ emitters, members of a 
protocluster at $z  = 4.1$, \citet{Overzier2008} derived a mean stellar mass of $\sim 
10^{8-9} M_{\sun}$ based on stacked $Ks$-band images, indicating that Ly$\alpha$ 
emitters in the field and in protoclusters at high redshift have similar masses. \citet{Kurk2004B} 
used near infrared magnitudes to derive the stellar masses of H$\alpha$ emitters 
found in the overdensity surrounding PKS~1138-262, a well known protocluster at $z = 2.16$, and
found that H$\alpha$ emitters are more massive than Ly$\alpha$ emitters, with a stellar mass 
$\sim 2 \times 10^{10} M_{\sun}$. The total stellar mass derived from both Ly$\alpha$ and H$\alpha$ emitters 
around PKS~1138-262 ($40$ sources) is $\sim10^{12} M_{\sun}$ \citep{Kurk2004B}. The mass function of 
galaxy clusters is in fact dominated by the evolved galaxy population, known to be 
rarer but much more massive than the narrow-band emitters \citep[e.g., for $Ks < 21.5$, 
$M_{stars} > 10^{11}M_{\sun}$;~][]{Kodama2007}.
If at $z\sim1.5$, the two objects with $Ks = 20$ found within $3\arcmin$ of 7C~1756+6520 
would each have a stellar mass of $5 \times 10^{11}  M_{\sun}$ and would therefore already have 
a mass equivalent to all the narrow-band emitters found near PKS 1138-262.
It is therefore essential to search for this population of red elliptical galaxies to fully understand
the earliest phases of cluster formation.

Our study makes use of wide-field optical and near-infrared cameras and permits 
the investigation of the spatial distribution of potential cluster members over a large 
area around the HzRG. Indeed, the small field of view of the previous generation of 
near-infrared instruments limited the study of large-scale structures, clusters and 
proto-clusters. Recently, \citet{Tanaka2007} presented a study of a large-scale structure 
around a galaxy cluster at $z=1.24$ with a possible large ($20$~Mpc) filamentary 
structure formed by the main cluster and four possible associated clumps of red 
galaxies, illustrating the necessity to look at galaxy clusters on larger scales than 
the cluster itself. The 7C~1756+6520 field presents several overdensities of red 
objects separated by several arcmin, as well as one nearby the HzRG. However, 
spectroscopic confirmation of these extended structures being associated and at high redshift is 
challenging due of the required large field of view on a multi-object spectrograph, and the
high redshift which places the main spectral features (emission lines for star-forming and breaks
for elliptical galaxies)  out of the optical bands. This will hopefully be achieved with the 
new generation of multi-object, near-infrared spectrographs (\eg MOIRCS on Subaru).

\begin{acknowledgements}

We are very grateful to S. Adam Stanford for useful discussions
and Tadayuki Kodama for having provided the models of red sequences presented in this paper. 
We thank Brigitte Rocca-Volmerange for her support of this project. We would also like to thank 
Andrea Grazian (and the GOODS-MUSIC team) and Ryan Quadri (and the MUSYC survey team) 
for useful emails exchanges on their online catalogs. This work is based in part on data products 
produced at the TERAPIX data center located at the Institut d'Astrophysique de Paris and generated 
from observations obtained at the Canada-France-Hawaii Telescope (CFHT) which is operated by 
the National Research Council of Canada, the Institut National des Sciences de 
l'Univers of the Centre National de la Recherche Scientifique of France, and the University of Hawaii.
It is also based on observations obtained at the Hale 200 inch telescope at Palomar Observatory and 
on observations made with the {\em Spitzer Space Telescope}, which is operated by the Jet Propulsion 
Laboratory, California Institute of Technology under a contract with NASA.
\end{acknowledgements}

\bibliographystyle{aa}
\bibliography{biblio}

\newpage

\clearpage

\begin{table*}
\caption{AGN candidates in 7C~1756+6520 field}
\label{AGN7C1756}
\begin{tabular}{c l l c c c c}
ID & R.A. & Dec. & [3.6] & [4.5] & [5.8] & [8.0] \\
\hline
HzRG	&	17:57:05.599	&	+65:19:53.86	&	20.01	&	19.87	&	20.08	&	19.74 \\
1  &	17:57:13.152	&	+65:17:06.43	&	19.41	&	19.51	&	19.64	&	19.46 \\
2  &	17:56:52.525	&	+65:16:56.67	&	20.12	&	19.67	&	19.78	&	18.93 \\
3  &	17:56:59.055	&	+65:17:54.96	&	19.82	&	19.48	&	19.43	&	19.20 \\
4  &	17:57:13.186	&	+65:19:08.40	&	19.77	&	19.15	&	18.51	&	18.45 \\
5  &	17:57:33.291	&	+65:20:25.55	&	18.77	&	18.76	&	18.95	&	18.10 \\
6  &	17:56:55.862	&	+65:19:06.56	&	18.32	&	17.82	&	17.56	&	17.16 \\
7  &	17:57:30.881	&	+65:21:21.77	&	18.98	&	18.57	&	17.98	&	17.64 \\
8  &	17:57:30.462	&	+65:21:21.75	&	19.12	&	18.58	&	17.97	&	17.50 \\
9  &	17:57:04.981	&	+65:19:51.00	&	20.02	&	19.61	&	19.69	&	19.34 \\
10 &	17:57:20.156	&	+65:21:56.79	&	20.16	&	19.99	&	19.63	&	19.33 \\
11 &	17:57:15.439	&	+65:21:56.02	&	19.13	&	19.17	&	19.73	&	19.62 \\
12 &	17:56:41.147	&	+65:22:59.14	&	17.48	&	17.24	&	18.42	&	18.36 \\
\hline                                             
\end{tabular}
\end{table*}


\begin{table*}
\caption{AGN candidates in 7C~1751+6809 field}
\label{AGN7C1751}
\begin{tabular}{cllcccc}
ID & R.A. & Dec. & [3.6] & [4.5] & [5.8] & [8.0] \\
\hline
HzRG	&	17:50:50.024	&	+68:08:26.47	&	19.68	&	19.63	&	20.78	&	19.99 \\
1	&	17:50:51.439	&	+68:06:06.59	&	18.52	&	18.23	&	18.56	&	18.02 \\
2	& 	17:51:24.879	&	+68:08:34.08	&	19.99	&	19.88	&	19.72	&	19.47 \\
3	&	17:50:39.182	&	+68:06:47.57	&	19.41	&	19.36	&	19.23	&	18.72 \\ 
4	&	17:50:34.238	&	+68:10:23.56	&	19.50	&	19.15	&	19.14	&	18.36 \\
5	&	17:50:51.762	&	+68:10:57.72	&	17.40	&	17.24	&	19.59	&	18.82 \\
\hline                                             
\end{tabular}
\end{table*}

\longtab{5}{
\begin{longtable}{c l l l c c c}
\caption{\label{7C1756pBzK} pBzK* galaxies in 7C~1756+6520 field} \\
ID & R.A. & Dec. & $B$ & $z$ & $J$ & $Ks$ \\
\hline
\endfirsthead
\caption{continued.} \\
ID & R.A. & Dec. & $B$ & $z$ & $J$ & $Ks$ \\
\hline
\endhead
\hline
\endfoot
       1 & 17:57:46.32 & 65:13:54.5 & 26.69\footnote[8]{Magnitudes are derived using SExtractor MAG\_AUTO, with the appropriate Galactic extinction correction applied. The color selection for p$BzK$* galaxies has been made from aperture magnitudes which give slightly different values.} & 21.55 & 20.60 & 20.11 \\
       2 & 17:58:01.68 & 65:15:14.7 & 26.78 & 22.39 & 20.93 & 20.15 \\
       3 & 17:55:50.88 & 65:16:20.6 & 24.63 & 22.91 & 21.16 & 20.33 \\
       4 & 17:58:10.56 & 65:16:56.6 & 25.65 & 22.71 & 21.59 & 20.57 \\
       5 & 17:57:19.19 & 65:19:31.1 & 25.94 & 22.72 & 21.57 & 20.67 \\
       6 & 17:58:26.40 & 65:19:35.4 & 26.45 & 24.22 & 22.78 & 21.77 \\
       7 & 17:57:12.48 & 65:19:45.1 & 26.65 & 22.63 & 21.37 & 20.18 \\ 
       8 & 17:56:30.72 & 65:20:52.4 & 25.20 & 23.24 & 22.43 & 20.99 \\
       9 & 17:56:16.80 & 65:21:10.4 & 23.49 & 21.90 & 21.58 & 20.55 \\
      10 & 17:56:48.48 & 65:21:22.0 & 25.01 & 23.14 & 22.04 & 20.97 \\
      11 & 17:56:52.56 & 65:22:59.5 & 26.88 & 23.70 & 22.55 & 20.71 \\
      12 & 17:57:01.68 & 65:16:50.2 & $>$27.1 & 24.18 & 22.91 & 22.03 \\
      13 & 17:57:12.48 & 65:19:41.9 & $>$27.1 & 23.04 & 21.84 & 20.63 \\
      14 & 17:57:20.40 & 65:11:51.0 & $>$27.1 & 23.03 & 22.08 & 20.92 \\
      15 & 17:57:12.48 & 65:12:10.5 & $>$27.1 & 23.68 & 22.63 & 21.30 \\
      16 & 17:57:53.76 & 65:12:45.7 & $>$27.1 & 23.29 & 21.95 & 21.09 \\
      17 & 17:57:33.12 & 65:13:13.4 & $>$27.1 & 24.11 & 22.58 & 21.81 \\
      18 & 17:57:28.80 & 65:13:28.2 & $>$27.1 & 23.45 & 21.99 & 21.18 \\
      19 & 17:57:11.04 & 65:13:46.9 & $>$27.1 & 22.25 & 21.12 & 20.44 \\
      20 & 17:57:40.08 & 65:14:08.9 & $>$27.1 & 23.28 & 22.30 & 21.33 \\
      21 & 17:57:41.76 & 65:14:17.9 & $>$27.1 & 24.25 & 22.97 & 21.71 \\
      22 & 17:57:26.40 & 65:14:59.3 & $>$27.1 & 23.88 & 22.45 & 21.74 \\
      23 & 17:57:52.56 & 65:15:16.9 & $>$27.1 & 23.76 & 22.93 & 21.57 \\
      24 & 17:57:40.32 & 65:15:23.4 & $>$27.1 & 23.13 & 21.53 & 20.51 \\
      25 & 17:57:16.56 & 65:15:43.9 & $>$27.1 & 22.79 & 21.46 & 20.49 \\
      26 & 17:58:06.48 & 65:15:47.2 & $>$27.1 & 23.13 & 21.96 & 20.74 \\
      27 & 17:58:13.92 & 65:15:51.1 & $>$27.1 & 22.08 & 20.95 & 20.11 \\
      28 & 17:56:55.68 & 65:16:00.5 & $>$27.1 & 23.81 & 22.13 & 21.20 \\
      29 & 17:56:25.44 & 65:16:36.1 & $>$27.1 & 23.37 & 22.80 & 21.57 \\
      30 & 17:58:19.68 & 65:16:50.5 & $>$27.1 & 23.39 & 21.94 & 21.20 \\
      31 & 17:57:44.40 & 65:17:14.3 & $>$27.1 & 23.80 & 23.56 & 21.62 \\
      32 & 17:57:40.08 & 65:17:17.5 & $>$27.1 & 23.32 & 21.84 & 21.10 \\
      33 & 17:57:33.84 & 65:17:20.8 & $>$27.1 & 23.22 & 22.70 & 21.58 \\
      34 & 17:56:43.92 & 65:17:50.6 & $>$27.1 & 23.43 & 22.99 & 21.69 \\
      35 & 17:58:29.04 & 65:18:25.2 & $>$27.1 & 22.34 & 20.83 & 20.33 \\
      36 & 17:56:32.64 & 65:18:32.8 & $>$27.1 & 23.79 & 22.25 & 21.46 \\
      37 & 17:56:48.96 & 65:18:50.8 & $>$27.1 & 23.70 & 22.65 & 21.32 \\
      38 & 17:56:52.32 & 65:18:52.2 & $>$27.1 & 22.55 & 21.21 & 20.31 \\
      39 & 17:57:28.32 & 65:19:24.6 & $>$27.1 & 21.91 & 20.82 & 19.93 \\
      40 & 17:57:13.92 & 65:19:44.1 & $>$27.1 & 23.16 & 21.82 & 20.79 \\
      41 & 17:57:22.80 & 65:19:45.5 & $>$27.1 & 23.17 & 22.39 & 21.46 \\
      42 & 17:57:58.08 & 65:19:44.4 & $>$27.1 & 23.91 & 22.33 & 21.40 \\
      43 & 17:57:05.04 & 65:19:54.5 & $>$27.1 & 23.47 & 22.48 & 21.31 \\
      44 & 17:56:42.72 & 65:20:27.2 & $>$27.1 & 23.11 & 21.91 & 20.93 \\
      45 & 17:57:21.36 & 65:20:52.1 & $>$27.1 & 22.50 & 21.38 & 20.30 \\
      46 & 17:57:52.08 & 65:21:07.9 & $>$27.1 & 23.70 & 22.17 & 21.48 \\
      47 & 17:56:55.44 & 65:21:15.5 & $>$27.1 & 22.97 & 21.78 & 20.80 \\
      48 & 17:57:40.08 & 65:21:26.6 & $>$27.1 & 24.75 & 22.31 & 22.10 \\
      49 & 17:58:16.56 & 65:21:24.1 & $>$27.1 & 23.18 & 21.74 & 20.92 \\
      50 & 17:56:19.44 & 65:21:25.9 & $>$27.1 & 22.05 & 20.52 & 19.87 \\
      51 & 17:56:35.28 & 65:21:32.8 & $>$27.1 & 23.85 & 22.83 & 21.33 \\
      52 & 17:56:50.16 & 65:21:56.9 & $>$27.1 & 23.88 & 22.62 & 21.54 \\
      53 & 17:56:50.16 & 65:22:05.9 & $>$27.1 & 23.16 & 22.22 & 20.94 \\
      54 & 17:58:36.96 & 65:22:09.5 & $>$27.1 & 23.63 & 22.46 & 21.20 \\
      55 & 17:56:12.00 & 65:22:19.9 & $>$27.1 & 23.45 & 21.74 & 21.25 \\
      56 & 17:56:25.20 & 65:22:23.9 & $>$27.1 & 24.13 & 22.69 & 21.42 \\
      57 & 17:55:31.68 & 65:22:36.8 & $>$27.1 & 21.72 & 21.03 & 20.14 \\
      58 & 17:55:55.44 & 65:23:17.5 & $>$27.1 & 22.99 & 22.05 & 21.32 \\
      59 & 17:57:33.36 & 65:23:21.8 & $>$27.1 & 24.34 & 23.25 & 21.83 \\
      60 & 17:56:54.48 & 65:23:35.1 & $>$27.1 & 22.41 & 21.25 & 20.23 \\
      61 & 17:58:07.69 & 65:23:50.6 & $>$27.1 & 21.89 & 20.63 & 19.79 \\
      62 & 17:57:11.04 & 65:23:56.8 & $>$27.1 & 23.40 & 22.51 & 21.20 \\
      63 & 17:57:25.20 & 65:23:58.2 & $>$27.1 & 23.45 & 22.86 & 21.76 \\
      64 & 17:56:49.44 & 65:24:21.2 & $>$27.1 & 23.41 & 22.36 & 21.13 \\
      65 & 17:57:45.84 & 65:24:34.6 & $>$27.1 & 23.62 & 21.77 & 21.39 \\
      66 & 17:55:42.72 & 65:24:56.9 & $>$27.1 & 22.80 & 21.30 & 20.43 \\
      67 & 17:58:14.88 & 65:25:05.2 & $>$27.1 & 22.59 & 21.22 & 20.32 \\
      68 & 17:57:00.00 & 65:25:12.4 & $>$27.1 & 23.54 & 22.16 & 21.73 \\
      69 & 17:57:54.72 & 65:25:25.7 & $>$27.1 & 24.14 & 22.42 & 21.53 \\
      70 & 17:55:33.60 & 65:25:26.4 & $>$27.1 & 22.99 & 21.66 & 20.56 \\
      71 & 17:55:57.84 & 65:25:37.6 & $>$27.1 & 22.63 & 21.51 & 20.59 \\
      72 & 17:57:39.12 & 65:25:40.1 & $>$27.1 & 23.87 & 23.07 & 21.62 \\
      73 & 17:55:57.84 & 65:25:40.8 & $>$27.1 & 23.65 & 22.21 & 21.36 \\
      74 & 17:57:16.08 & 65:26:01.7 & $>$27.1 & 23.72 & 22.19 & 21.36 \\
      75 & 17:55:55.44 & 65:26:12.8 & $>$27.1 & 23.79 & 22.19 & 21.43 \\
      76 & 17:55:39.60 & 65:26:21.8 & $>$27.1 & 23.32 & 21.52 & 20.83 \\
      77 & 17:57:05.28 & 65:27:21.6 & $>$27.1 & 23.99 & 22.63 & 21.80 \\
      78 & 17:55:29.76 & 65:27:33.8 & $>$27.1 & 22.55 & 21.14 & 20.41 \\
      79 & 17:55:46.56 & 65:27:32.8 & $>$27.1 & 23.65 & 22.54 & 21.64 \\
      80 & 17:57:28.80 & 65:27:34.2 & $>$27.1 & 23.20 & 22.42 & 20.72 \\
      81 & 17:56:58.80 & 65:27:39.2 & $>$27.1 & 23.06 & 22.19 & 20.44 \\
      82 & 17:55:59.28 & 65:27:41.8 & $>$27.1 & 22.63 & 22.03 & 20.78 \\
      83 & 17:57:35.04 & 65:27:38.9 & $>$27.1 & 23.74 & 22.16 & 21.30 \\
      84 & 17:55:57.12 & 65:27:48.2 & $>$27.1 & 22.30 & 20.81 & 20.18 \\
      85 & 17:55:57.84 & 65:27:49.0 & $>$27.1 & 24.36 & 22.55 & 21.85 \\
      86 & 17:56:49.68 & 65:27:56.2 & $>$27.1 & 22.60 & 21.37 & 20.63 \\
      87 & 17:56:42.48 & 65:28:17.4 & $>$27.1 & 24.19 & 22.86 & 21.62 \\
      88 & 17:56:47.28 & 65:28:36.5 & $>$27.1 & 23.48 & 22.06 & 21.32 \\
      89 & 17:57:06.00 & 65:29:08.5 & $>$27.1 & 23.90 & 23.24 & 21.40 \\
      90 & 17:57:34.32 & 65:11:39.1 & $>$27.1 & 22.33 & 20.73 & 19.78 \\
      91 & 17:57:10.08 & 65:12:55.4 & $>$27.1 & 22.81 & 21.83 & 20.89 \\
      92 & 17:57:09.84 & 65:12:59.4 & $>$27.1 & 23.63 & 22.73 & 21.49 \\
      93 & 17:57:36.48 & 65:13:50.2 & $>$27.1 & 23.85 & 22.20 & 21.32 \\
      94 & 17:58:14.64 & 65:13:46.6 & $>$27.1 & 24.21 & 22.10 & 21.00 \\
      95 & 17:56:41.04 & 65:13:55.9 & $>$27.1 & 23.91 & 21.91 & 20.98 \\
      96 & 17:57:46.56 & 65:13:56.3 & $>$27.1 & 22.46 & 21.05 & 20.25 \\
      97 & 17:57:34.56 & 65:14:06.4 & $>$27.1 & 23.83 & 22.59 & 21.18 \\
      98 & 17:56:25.68 & 65:14:45.6 & $>$27.1 & 23.93 & 21.77 & 21.30 \\
      99 & 17:57:59.28 & 65:15:07.9 & $>$27.1 & 23.44 & 22.46 & 21.03 \\
     100 & 17:57:19.44 & 65:15:24.5 & $>$27.1 & 23.54 & 22.48 & 21.45 \\
     101 & 17:57:53.76 & 65:15:33.1 & $>$27.1 & 23.45 & 22.21 & 21.02 \\
     102 & 17:56:53.04 & 65:15:38.9 & $>$27.1 & 24.05 & 22.25 & 21.31 \\
     103 & 17:58:14.88 & 65:15:55.5 & $>$27.1 & 23.18 & 21.64 & 20.29 \\
     104 & 17:57:26.16 & 65:17:08.5 & $>$27.1 & 23.07 & 21.68 & 20.60 \\
     105 & 17:57:51.36 & 65:17:17.9 & $>$27.1 & 22.78 & 21.52 & 20.46 \\
     106 & 17:57:21.12 & 65:17:24.7 & $>$27.1 & 21.95 & 21.02 & 19.94 \\
     107 & 17:57:59.76 & 65:18:01.8 & $>$27.1 & 23.70 & 22.07 & 21.16 \\
     108 & 17:55:27.12 & 65:18:25.6 & $>$27.1 & 23.61 & 21.80 & 21.03 \\
     109 & 17:56:33.60 & 65:18:31.0 & $>$27.1 & 23.94 & 22.12 & 21.18 \\
     110 & 17:58:21.12 & 65:19:13.8 & $>$27.1 & 22.86 & 21.52 & 20.31 \\
     111 & 17:57:29.28 & 65:19:33.6 & $>$27.1 & 22.70 & 21.30 & 20.16 \\
     112 & 17:57:10.08 & 65:20:39.5 & $>$27.1 & 23.78 & 22.58 & 21.48 \\
     113 & 17:56:53.76 & 65:21:07.5 & $>$27.1 & 23.76 & 22.40 & 21.20 \\
     114 & 17:58:22.80 & 65:21:31.3 & $>$27.1 & 22.41 & 20.66 & 19.95 \\
     115 & 17:56:55.20 & 65:21:54.0 & $>$27.1 & 23.65 & 21.54 & 20.59 \\
     116 & 17:56:45.84 & 65:22:27.5 & $>$27.1 & 23.42 & 21.49 & 20.71 \\
     117 & 17:57:02.40 & 65:22:40.4 & $>$27.1 & 23.42 & 22.80 & 20.94 \\
     118 & 17:57:12.48 & 65:23:25.8 & $>$27.1 & 23.42 & 21.83 & 20.58 \\
     119 & 17:56:46.80 & 65:24:14.8 & $>$27.1 & 23.29 & 21.48 & 20.69 \\
     120 & 17:58:11.04 & 65:24:22.7 & $>$27.1 & 23.72 & 22.77 & 21.36 \\
     121 & 17:56:18.96 & 65:24:50.4 & $>$27.1 & 23.91 & 22.89 & 21.34 \\
     122 & 17:57:58.56 & 65:24:52.9 & $>$27.1 & 23.55 & 21.78 & 20.96 \\
     123 & 17:56:16.56 & 65:25:41.9 & $>$27.1 & 23.79 & 22.87 & 21.47 \\
     124 & 17:55:55.20 & 65:26:33.3 & $>$27.1 & 22.44 & 21.36 & 20.48 \\
     125 & 17:56:50.16 & 65:27:05.0 & $>$27.1 & 23.71 & 22.70 & 21.23 \\
     126 & 17:56:33.36 & 65:27:03.2 & $>$27.1 & 22.20 & 21.04 & 20.11 \\
     127 & 17:56:15.12 & 65:27:56.5 & $>$27.1 & 22.93 & 21.63 & 20.52 \\
     128 & 17:56:13.20 & 65:28:44.8 & $>$27.1 & 22.70 & 21.20 & 20.19 \\
     129 & 17:57:48.72 & 65:29:17.9 & $>$27.1 & 22.98 & 21.43 & 20.01 \\                                        
\end{longtable}}

\longtab{6}{
\begin{longtable}{c l l l c c c}
\caption{\label{7C1751pBzK} pBzK* galaxies in 7C~1751+6809 field} \\
ID & R.A. & Dec. & $B$ & $z$ & $J$ & $Ks$ \\
\hline
\endfirsthead
\caption{continued.} \\
ID & R.A. & Dec. & $B$ & $z$ & $J$ & $Ks$ \\
\hline
\endhead
\hline
\endfoot
1	&	17:51:44.88	&	67:59:43.1	&	25.24\footnote[9]{See Table 5, note 8} 	&	23.73	&	22.58	&	21.05	\\
2	&	17:51:07.92	&	68:04:10.9	&	25.82	&	22.36	&	21.12	&	20.16	\\
3	&	17:49:31.44	&	68:11:44.5	&	24.74	&	22.67	&	21.59	&	20.47	\\
4	&	17:51:31.68	&	68:11:42.7	&	25.17	&	24.10	&	22.51	&	21.30	\\
5	&	17:49:57.37	&	68:13:54.5	&	24.40	&	21.77	&	20.98	&	19.81	\\
6	&	17:50:52.08	&	68:15:00.7	&	25.35	&	22.37	&	60.86	&	20.07	\\
7	&	17:50:05.76	&	68:16:30.4	&	24.22	&	22.96	&	21.71	&	20.70	\\
8	&	17:51:49.92	&	68:17:51.4	&	24.54	&	23.33	&	21.86	&	20.53	\\
9	&	17:50:09.12	&	67:58:32.2	&	$>$27.1	&	23.93	&	22.96	&	21.10	\\
10	&	17:51:38.65	&	67:58:41.2	&	$>$27.1	&	24.51	&	22.31	&	21.43	\\
11	&	17:49:57.12	&	67:59:24.3	&	$>$27.1	&	23.94	&	22.39	&	21.33	\\
12	&	17:51:50.64	&	68:01:09.8	&	$>$27.1	&	22.62	&	21.36	&	20.57	\\
13	&	17:50:32.16	&	68:01:21.4	&	$>$27.1	&	24.21	&	22.35	&	21.29	\\
14	&	17:51:06.48	&	68:01:28.9	&	$>$27.1	&	22.77	&	21.67	&	20.85	\\
15	&	17:51:24.96	&	68:02:54.2	&	$>$27.1	&	22.22	&	21.31	&	20.07	\\
16	&	17:51:43.20	&	68:03:38.5	&	$>$27.1	&	22.85	&	21.61	&	20.90	\\
17	&	17:49:00.00	&	68:04:08.4	&	$>$27.1	&	22.24	&	21.48	&	20.51	\\
18	&	17:51:19.44	&	68:04:14.9	&	$>$27.1	&	24.07	&	21.75	&	21.59	\\
19	&	17:52:10.56	&	68:04:24.6	&	$>$27.1	&	24.07	&	21.95	&	21.88	\\
20	&	17:49:42.48	&	68:04:32.5	&	$>$27.1	&	24.74	&	22.41	&	21.83	\\
21	&	17:52:21.36	&	68:04:39.0	&	$>$27.1	&	23.67	&	22.08	&	20.80	\\
22	&	17:50:19.44	&	68:05:04.9	&	$>$27.1	&	24.23	&	22.71	&	21.37	\\
23	&	17:48:53.52	&	68:05:08.5	&	$>$27.1	&	23.30	&	22.03	&	21.28	\\
24	&	17:49:42.00	&	68:05:11.1	&	$>$27.1	&	22.93	&	21.42	&	20.66	\\
25	&	17:49:38.16	&	68:05:25.4	&	$>$27.1	&	23.86	&	21.96	&	21.18	\\
26	&	17:50:35.52	&	68:06:08.6	&	$>$27.1	&	23.52	&	22.14	&	21.17	\\
27	&	17:51:19.44	&	68:06:30.6	&	$>$27.1	&	23.66	&	22.15	&	21.00	\\
28	&	17:51:52.80	&	68:07:06.6	&	$>$27.1	&	21.75	&	20.94	&	19.63	\\
29	&	17:51:04.56	&	68:07:06.6	&	$>$27.1	&	23.25	&	21.73	&	20.76	\\
30	&	17:49:10.56	&	68:07:18.1	&	$>$27.1	&	23.16	&	21.99	&	20.91	\\
31	&	17:52:22.56	&	68:07:51.6	&	$>$27.1	&	24.63	&	22.13	&	20.98	\\
32	&	17:49:01.68	&	68:07:58.1	&	$>$27.1	&	24.03	&	22.04	&	21.35	\\
33	&	17:50:49.92	&	68:08:26.2	&	$>$27.1	&	22.40	&	21.11	&	20.04	\\
34	&	17:51:13.93	&	68:08:27.6	&	$>$27.1	&	22.67	&	20.83	&	20.18	\\
35	&	17:48:52.32	&	68:08:26.2	&	$>$27.1	&	24.12	&	22.16	&	21.11	\\
36	&	17:51:13.44	&	68:08:29.8	&	$>$27.1	&	23.53	&	21.78	&	20.94	\\
37	&	17:51:24.96	&	68:08:33.7	&	$>$27.1	&	22.69	&	21.51	&	20.70	\\
38	&	17:49:14.64	&	68:09:10.1	&	$>$27.1	&	22.80	&	21.14	&	20.29	\\
39	&	17:49:08.40	&	68:09:12.6	&	$>$27.1	&	22.95	&	21.62	&	20.38	\\
40	&	17:48:50.64	&	68:09:53.3	&	$>$27.1	&	23.23	&	21.20	&	20.59	\\
41	&	17:49:46.80	&	68:11:01.7	&	$>$27.1	&	23.56	&	21.75	&	21.12	\\
42	&	17:50:17.52	&	68:11:20.1	&	$>$27.1	&	23.80	&	21.95	&	21.20	\\
43	&	17:48:48.23	&	68:11:30.8	&	$>$27.1	&	20.56	&	19.37	&	18.57	\\
44	&	17:48:53.52	&	68:11:47.1	&	$>$27.1	&	23.43	&	21.59	&	20.93	\\
45	&	17:49:08.87	&	68:11:48.5	&	$>$27.1	&	22.81	&	21.63	&	20.77	\\
46	&	17:51:53.76	&	68:12:02.5	&	$>$27.1	&	23.88	&	60.86	&	21.69	\\
47	&	17:49:8.16	&	68:12:18.4	&	$>$27.1	&	22.87	&	21.24	&	20.56	\\
48	&	17:49:6.24	&	68:12:28.1	&	$>$27.1	&	22.25	&	20.96	&	20.14	\\
49	&	17:49:7.92	&	68:12:38.9	&	$>$27.1	&	23.82	&	21.41	&	20.92	\\
50	&	17:50:18.72	&	68:13:32.2	&	$>$27.1	&	23.49	&	22.18	&	21.03	\\
51	&	17:51:42.24	&	68:13:40.1	&	$>$27.1	&	23.91	&	21.51	&	20.99	\\
52	&	17:49:40.32	&	68:14:17.2	&	$>$27.1	&	23.94	&	22.02	&	21.41	\\
53	&	17:50:53.52	&	68:14:39.5	&	$>$27.1	&	21.91	&	20.97	&	20.17	\\
54	&	17:49:59.04	&	68:14:35.2	&	$>$27.1	&	22.70	&	21.21	&	20.49	\\
55	&	17:49:07.92	&	68:14:43.8	&	$>$27.1	&	24.21	&	22.36	&	21.04	\\
56	&	17:52:06.72	&	68:15:00.4	&	$>$27.1	&	23.30	&	21.11	&	20.91	\\
57	&	17:51:12.00	&	68:15:07.6	&	$>$27.1	&	22.94	&	20.98	&	20.43	\\
58	&	17:52:10.32	&	68:15:21.2	&	$>$27.1	&	21.95	&	20.51	&	19.74	\\
59	&	17:49:55.68	&	68:16:18.5	&	$>$27.1	&	23.12	&	21.69	&	20.88	\\
60	&	17:50:04.80	&	68:16:25.7	&	$>$27.1	&	22.90	&	21.45	&	20.63	\\
61	&	17:49:57.37	&	68:16:37.9	&	$>$27.1	&	22.92	&	21.47	&	20.60	\\
62	&	17:50:43.20	&	68:16:52.3	&	$>$27.1	&	21.61	&	20.86	&	19.92	\\
63	&	17:50:57.84	&	68:16:50.5	&	$>$27.1	&	24.32	&	22.01	&	21.12	\\
64	&	17:51:59.76	&	68:18:07.6	&	$>$27.1	&	22.84	&	21.36	&	20.59	\\
65	&	17:49:59.76	&	67:58:59.5	&	$>$27.1	&	24.38	&	22.23	&	21.42	\\
66	&	17:50:49.92	&	68:01:34.3	&	$>$27.1	&	22.91	&	21.27	&	20.39	\\
67	&	17:51:24.72	&	68:01:41.9	&	$>$27.1	&	23.38	&	21.41	&	20.17	\\
68	&	17:51:19.44	&	68:03:00.0	&	$>$27.1	&	23.79	&	21.87	&	20.74	\\
69	&	17:51:30.48	&	68:03:27.4	&	$>$27.1	&	23.35	&	21.46	&	20.57	\\
70	&	17:51:40.32	&	68:03:37.1	&	$>$27.1	&	23.81	&	22.06	&	20.96	\\
71	&	17:49:20.64	&	68:04:45.8	&	$>$27.1	&	23.38	&	21.49	&	20.50	\\
72	&	17:50:47.52	&	68:05:13.9	&	$>$27.1	&	23.70	&	22.34	&	20.82	\\
73	&	17:52:06.96	&	68:06:02.9	&	$>$27.1	&	18.65	&	60.86	&	16.64	\\
74	&	17:52:10.08	&	68:06:22.0	&	$>$27.1	&	18.43	&	60.86	&	16.13	\\
75	&	17:48:54.00	&	68:07:07.3	&	$>$27.1	&	19.29	&	60.86	&	16.74	\\
76	&	17:50:38.40	&	68:07:05.9	&	$>$27.1	&	24.09	&	21.95	&	21.26	\\
77	&	17:49:56.88	&	68:07:07.7	&	$>$27.1	&	24.29	&	22.79	&	21.27	\\
78	&	17:52:05.75	&	68:07:12.4	&	$>$27.1	&	24.28	&	22.03	&	20.88	\\
79	&	17:49:41.04	&	68:07:18.1	&	$>$27.1	&	23.47	&	21.69	&	20.97	\\
80	&	17:50:23.28	&	68:07:29.3	&	$>$27.1	&	23.53	&	21.94	&	20.76	\\
81	&	17:49:43.92	&	68:07:54.1	&	$>$27.1	&	23.74	&	21.90	&	21.21	\\
82	&	17:51:15.60	&	68:08:19.0	&	$>$27.1	&	24.08	&	22.32	&	21.41	\\
83	&	17:50:48.96	&	68:08:24.7	&	$>$27.1	&	23.21	&	20.80	&	20.48	\\
84	&	17:51:40.56	&	68:08:47.0	&	$>$27.1	&	23.41	&	21.79	&	20.77	\\
85	&	17:49:06.49	&	68:08:57.1	&	$>$27.1	&	23.07	&	21.03	&	20.19	\\
86	&	17:51:29.51	&	68:09:00.7	&	$>$27.1	&	23.48	&	22.13	&	20.65	\\
87	&	17:50:32.64	&	68:09:07.9	&	$>$27.1	&	24.08	&	21.79	&	20.92	\\
88	&	17:49:22.09	&	68:10:08.0	&	$>$27.1	&	23.01	&	21.30	&	20.35	\\
89	&	17:51:35.04	&	68:10:12.4	&	$>$27.1	&	23.73	&	21.89	&	21.38	\\
90	&	17:49:16.80	&	68:11:15.0	&	$>$27.1	&	22.42	&	21.14	&	20.13	\\
91	&	17:52:28.56	&	68:12:09.7	&	$>$27.1	&	24.15	&	21.54	&	20.70	\\
92	&	17:51:29.04	&	68:12:57.6	&	$>$27.1	&	23.37	&	21.09	&	20.84	\\
93	&	17:49:36.48	&	68:13:08.8	&	$>$27.1	&	23.81	&	21.36	&	20.94	\\
94	&	17:48:56.16	&	68:13:18.8	&	$>$27.1	&	22.75	&	21.14	&	19.94	\\
95	&	17:48:58.08	&	68:13:43.7	&	$>$27.1	&	23.67	&	22.19	&	20.83	\\
96	&	17:51:21.36	&	68:13:49.1	&	$>$27.1	&	23.07	&	21.53	&	20.64	\\
97	&	17:50:05.52	&	68:14:05.6	&	$>$27.1	&	22.95	&	21.00	&	20.57	\\
98	&	17:49:06.96	&	68:14:19.3	&	$>$27.1	&	23.11	&	21.22	&	20.36	\\
99	&	17:50:05.04	&	68:14:22.6	&	$>$27.1	&	23.35	&	60.86	&	20.56	\\
100	&	17:49:13.92	&	68:14:31.6	&	$>$27.1	&	22.52	&	60.86	&	19.85	\\
101	&	17:51:16.08	&	68:14:49.2	&	$>$27.1	&	23.81	&	22.27	&	20.93	\\
102	&	17:49:55.92	&	68:15:12.2	&	$>$27.1	&	23.70	&	21.68	&	20.88	\\
103	&	17:51:12.48	&	68:15:11.5	&	$>$27.1	&	23.55	&	21.50	&	20.79	\\
104	&	17:51:24.00	&	68:15:33.8	&	$>$27.1	&	23.80	&	21.69	&	21.21	\\
105	&	17:49:24.72	&	68:15:47.2	&	$>$27.1	&	23.93	&	22.54	&	21.30	\\
106	&	17:51:49.92	&	68:17:51.4	&	$>$27.1	&	23.53	&	60.86	&	20.53	\\
\end{longtable}}

\end{document}